\documentclass[11pt]{article}

\usepackage{url} 
\usepackage[T1]{fontenc}
\usepackage{amsmath,amssymb}
\usepackage{bm}
\usepackage{subcaption}
\usepackage{lineno}
\usepackage{soul}
\usepackage{hhline}
\usepackage{comment}
\usepackage{natbib}
\usepackage[dvipsnames]{xcolor}
\usepackage{gensymb}
\usepackage{hyperref}
\usepackage{aas_macros}
\usepackage[left=2.5cm, right=2.5cm, top=2.5cm, bottom=2.5cm]{geometry}
\usepackage{graphicx}

\newcommand{\kmps}{\text{km s}^{-1}}          
\newcommand{\At}{A_\text{t}}
\newcommand{\Ata}{A_\text{t1}}
\newcommand{\Atb}{A_\text{t2}}
\newcommand{\tnl}{\tau_\text{nl}}

\begin{document}

\title{\textbf{Dynamical Age of Alfv\'enic Turbulence in the Solar Wind}}

\author{\textbf{Rohit Chhiber}$^{1,2}$\thanks{\url{rohit.chhiber@nasa.gov},\url{rohitc@udel.edu}}~~, \textbf{Yanwen Wang}$^{3}$,
\textbf{Arcadi V. Usmanov}$^{1,2}$,
and \textbf{William H. Matthaeus}$^{1}$
\\ \\
$^{1}$Department of Physics and Astronomy, University of Delaware, Newark, DE 19716, USA\\
$^{2}$Heliophysics Science Division, NASA Goddard Space Flight Center, Greenbelt, MD 20771, USA\\
$^{3}$Department of Physics, University of Maryland, College Park, MD 20742, USA
}

\maketitle

\begin{abstract}
An evolving turbulent flow such as the solar wind can be meaningfully characterized by its ``\textit{turbulence age}'' -- an estimate of the number of nonlinear times that have elapsed during a plasma parcel's propagation from the Sun to a given point in space. Recent observations of the near-Sun solar wind by the \textit{Parker Solar Probe} (\textit{PSP}) indicate high correlation between velocity and magnetic fluctuations (i.e., cross helicity, \(\sigma_c\)), which is known to impede development of magnetohydrodynamic (MHD) turbulence. Here we propose a new formulation of the turbulence age (\(A_\text{t}\)) of the solar wind that explicitly accounts for the Alfv\'enic nature of the fluctuations in the inner heliosphere. \(A_\text{t}\) is then evaluated for slow and fast wind streams using a variety of data sources -- observations from the \textit{PSP, Advanced Composition Explorer}, and \textit{Voyager} missions, and a global solar wind simulation that includes turbulence transport. Compared to the formulation employed in previous work that neglected Alfv\'enicity, the present approach yields smaller values of \(A_\text{t}\) in medium-to-high \(\sigma_c\) solar wind; similar turbulence ages are then obtained for slow and fast wind in the ecliptic. The radial evolution of \(A_\text{t}\) between heliocentric distances of  \(r\sim 0.2\) to 40 AU is examined. The rate of increase of \(A_\text{t}\) is found to decrease until \(\sim 5\) AU, indicating a gradual slowing of the \textit{in situ} development of turbulence in the inner heliosphere. Beyond \(\sim 5\) AU this rate begins to increase, likely due to turbulence driving by pick-up ions. This paper highlights the important role of \(\sigma_c\) in modulating MHD turbulence, and the results will aid in further interpretations of observations of the radial evolution of various turbulence parameters in the solar wind.
\end{abstract}

\section{Introduction}

Large-amplitude fluctuations in the velocity and magnetic fields of the solar wind have been recorded since the dawn of the space age \citep{belcher1971JGR}. The associated turbulence cascade is recognized as an important heliospheric phenomenon that can play a fundamental role in heating and accelerating the plasma \citep{matthaeus2011SSR} and transporting solar energetic particles (SEPs) and cosmic rays \citep{engelbrecht2022ssr}. While many theoretical approaches employed in the study of heliospheric turbulence can trace their roots to hydrodynamics, the interaction of the plasma with electromagnetic fields introduces substantial modifications \citep{zhou2004RMP}. A key aspect is the Alfv\'en-wave like nature of the observed turbulence, especially in regard to the cross helicity  \citep[\(\sigma_c\), the correlation of velocity and magnetic fluctuations][]{bruno2013LRSP}. 

This Alfv\'enic character is related to the solar source, where convective motions below the photosphere can induce fluctuations in the magnetic ``footpoints'' of the solar wind, thus launching Alfv\'en waves into the atmosphere \citep{DePontieu2007Sci}. Observations of fast wind (especially at high heliolatitudes) indicate that the embedded fluctuations have large \(\sigma_c\) even at several AU from the Sun, while the slow ecliptic wind generally has low \(\sigma_c\) at 1 AU \citep{bruno2013LRSP}. Recent observations by the \textit{Parker Solar Probe} (\textit{PSP}) indicate that even the slow wind has large \(\sigma_c\) near the Sun \citep[helioradius \(r\lesssim0.2\) AU;][]{chen2020ApJS}. Turbulence dynamics in this crucial region, where the wind is heated and accelerated and SEPs injected into the heliosphere \citep{Raouafi2023SSR}, are therefore likely to be significantly influenced by the large \(\sigma_c\). Theoretical work on magnetohydrodynamic (MHD) turbulence has shown that large \(\sigma_c\) can suppress the efficacy of the nonlinear interactions that produce turbulent cascade \citep{kraichnan1965PoF,zhou2004RMP}. Indeed, recent studies have characterized near-Sun turbulence as less developed compared to  1 AU (and beyond), in terms of standard diagnostics like spectral slope and multifractal statistics \citep{alberti2020psp,Telloni2021ApJ}. 

The concept of a \textit{turbulence age} (\(A_\text{t}\)) is of utility in the interpretation of such observations of evolving turbulent flows. It represents the level of turbulence development in terms of the number of nonlinear times that have elapsed during the passage of a flow between two points in space and/or time, thus defining a natural, intrinsic ``clock'' for the dynamics. Arbitrary time intervals can then be expressed in units of \(A_\text{t}\), and the relative importance of turbulence compared to other relevant processes can be evaluated by comparison of characteristic timescales.\footnote{\(\At\) is analogous to ``collisional age'' -- the number of Coulomb collisions experienced by a plasma parcel to an  observation point \citep{chhiber2016solar} -- widely used to organize solar wind based on degree of departure from thermal equilibrium: collisionally ``young'' wind exhibits stronger kinetic (non-MHD) effects like pronounced anisotropy in ion temperatures, and differential heating and acceleration of plasma species \citep{Verscharen2019LRSP}.} 

In the context of the solar wind, the turbulence age concept was introduced by \cite{matthaeus1998JGR} (henceforth M1998), who proposed and evaluated various formulations that employed both modeling and observations. They found that the solar wind experiences around 40  ``eddy'' turnover times between 1 and 40 AU, with a gradual ``slowing'' of the turbulence as it ages. The concept has since then been employed in a number of studies, including a recent analysis of \textit{PSP} data that investigated whether the observed variability of the magnetic field's spectral index was organized by by \(\At\) \citep{McIntyre2023ApJ}. However, M1998 neglected Alfv\'enicity in their formulations and left consideration of non-zero \(\sigma_c\) for future work. It is the intent of this paper to modify the M1998 formulation to account for cross helicity effects, compare the two formulations, and evaluate the turbulence age of the solar wind in an expanded domain that includes the young solar wind (\(r<1\) AU) and polar regions. Observations from the \textit{PSP,~Advanced Composition Explorer} (\textit{ACE}), and \textit{Voyager} missions will be employed, along with data from a global solar wind simulation that includes turbulence transport. 

\section{Turbulence age of the solar wind and its modulation by cross helicity}\label{sec:model}

M1998 define the dimensionless age of turbulence as 
\begin{equation}
    A_\text{t} = \int_{t_1}^{t_2} \frac{dt'}{\tnl(t')}, \label{eq:At}
\end{equation}
where \(t_{1,2}\) are two reference times, and the energy decay equation
\begin{equation}
    \frac{dE}{dt}=-\frac{E}{\tnl} \label{eq:vK-H_hydro}
\end{equation}
for the fluid-scale energy per unit mass \(E\) can serve to define the characteristic nonlinear turbulence timescale \(\tnl\). With a change of variable from time to radial distance in the solar wind, \(dt'=dr/U\), where \(U(r)\) is the solar wind's radial speed,
\begin{equation}
    A_\text{t} = \int_{r_1}^{r_2} \frac{dr}{U(r)\tnl(r)}.
\end{equation}
%
M1998 propose the use of a Taylor-Karman decay phenomenology \citep{Batchelor1953book} in which the turbulence timescale becomes the ``eddy turnover time'': \(\tnl=\lambda(t)/u(t)\), where  \(\lambda\) is the energy-containing (or correlation) scale, and \(u\sim \sqrt{2 E(t)}\) is a characteristic turbulence speed. The energy decay model can be written as \(\dot{u^2}=-u^3/\lambda\) \citep{karman1938prsl} for the hydrodynamic case.\footnote{We have not displayed the \(\mathcal{O}(1)\) proportionality constants that appear in these equations \citep[e.g.,][]{hossain1995PhFl} since they do not have bearing on our considerations.}

In the case of incompressible MHD turbulence, the transport equations are often written in terms of the Els\"asser variables \(\bm{z}_\pm=\bm{u}\pm \bm{b}\) \citep{elsasser1950PhRv}, for velocity (\(\bm{u}\)) and magnetic fluctuations (\(\bm{b}\)), where it is understood that the latter are in Alfv\'en units (\(\bm{b}\rightarrow \bm{b}/\sqrt{4\pi\rho}\), where \(\rho\) is mass density). Normalized cross helicity is defined as \(\sigma_c\equiv 2\langle\bm{u}\cdot\bm{b}\rangle Z^{-2}\), where \(\langle\cdot\rangle\) indicates an average over a suitable domain, and \(Z^2 = \langle u^2+b^2\rangle \) is twice the total fluctuation energy per unit mass. It is also conveniently expressed as \(\sigma_c\equiv(Z_+^2-Z_-^2)/(Z_+^2+Z_-^2)\), where the separate Els\"asser energies are \(Z^2_\pm\equiv \langle z_\pm^2\rangle\). When either of \(\bm{z}_\pm\) is zero, nonlinear interactions vanish and the remaining \(\bm{z}\) corresponds to a unidirectional, propagating Alfv\'en wave, with \(\sigma_c=\pm1\). When both \(\bm{z}_\pm\) are non-zero they interact nonlinearly, producing turbulence. For ``balanced'' turbulence \(Z_+=Z_-\equiv Z\), the corresponding zero-\(\sigma_c\) MHD timescale is \(\tnl=\lambda/Z\) \citep{kraichnan1965PoF,Dobrowolny1980PRL}, and the energy decay model is \(\dot{Z^2}=-Z^3/\lambda\) \citep{hossain1995PhFl}. Then \(A_\text{t}\) can be directly computed as 
\begin{equation}
    A_\text{t1} \equiv \int_{r1}^{r2} \frac{dr}{U(r)} \frac{Z(r)}{\lambda(r)}. \label{eq:At1}
\end{equation}
The above formulation was employed in M1998; we denote it \(\Ata\).

\begin{figure}
    \centering
       \includegraphics[width=.4\textwidth]{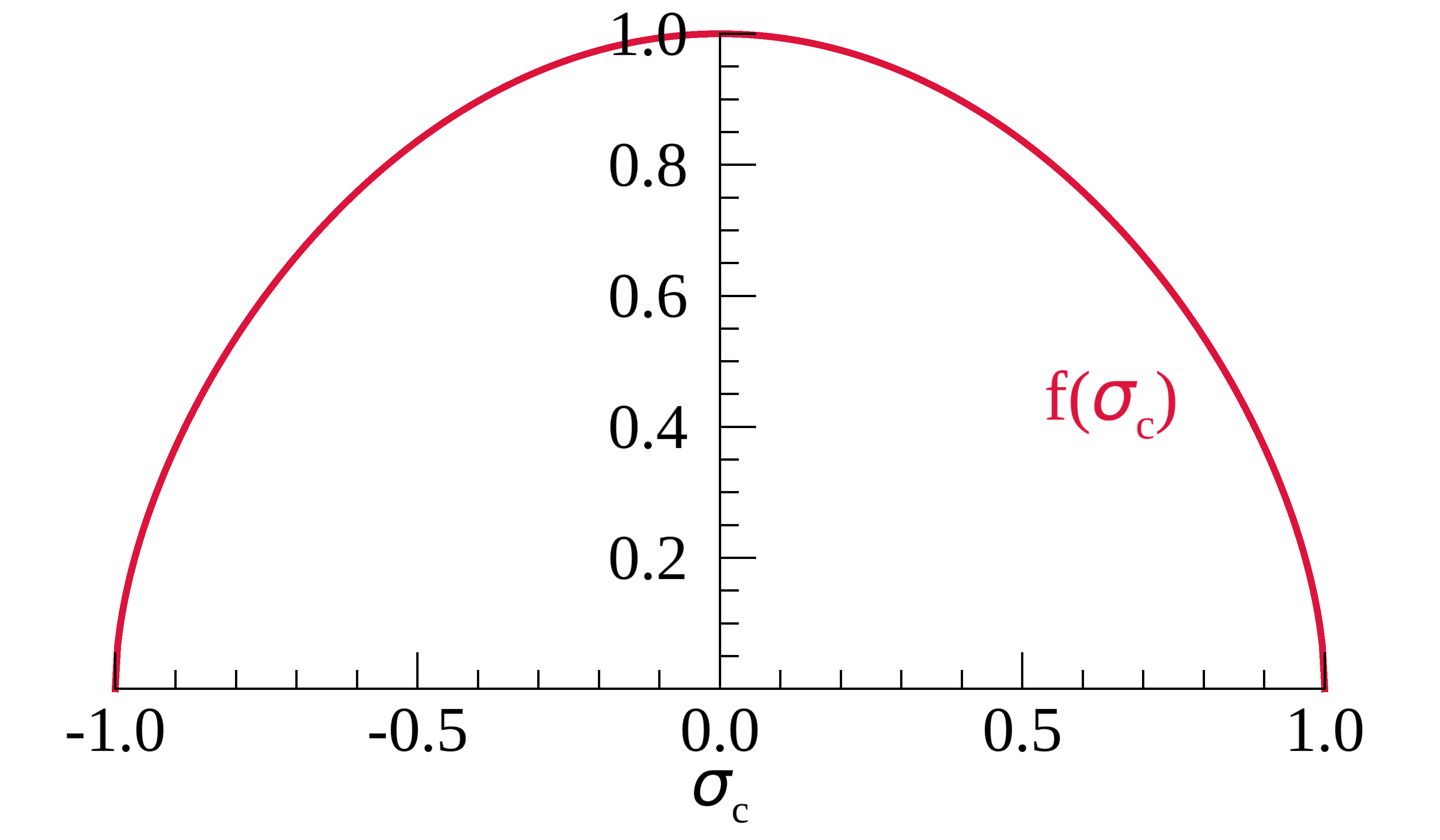}
    \caption{The function \(f(\sigma_c)\) [Eq. \eqref{eq:f_sigc}].}
    \label{fig:f}
\end{figure}

For non-zero \(\sigma_c\) the energy decay rate can be written as follows, based on analysis of the MHD equations \citep{Dobrowolny1980PRL,hossain1995PhFl}:
\begin{equation}
    \frac{dZ_\pm^2}{dt} = - \frac{Z_\pm^2 Z_\mp}{\lambda},
\end{equation}
where we do not distinguish between the similarity lengthscales \(\lambda_\pm\) \citep[see][]{breech2008turbulence,chhiber2021ApJ_psp}. Use of the identities \(Z^2_\pm=(1\pm\sigma_c)Z^2\) yields \citep{matthaeus2004grl}:
\begin{equation}
    \frac{dZ^2}{dt} = - f(\sigma_c) \frac{Z^3}{\lambda} = - f(\sigma_c) \frac{Z^2}{\tnl}, \label{eq:vKH_MHD}
\end{equation}
where
\begin{equation}
    f(\sigma_c)=\frac{(1-\sigma_c^2)^{1/2}}{2} \left[(1+\sigma_c)^{1/2} + (1-\sigma_c)^{1/2} \right]. \label{eq:f_sigc}
\end{equation}
The behavior of \(f(\sigma_c)\) as a function of cross helicity is shown in Figure \ref{fig:f}.

Comparing Eq. \eqref{eq:vKH_MHD} with Eq. \eqref{eq:vK-H_hydro}, we can define an effective turbulence timescale \(\tnl'=\tnl /f(\sigma_c)\). This has the attractive properties: when \(\sigma_c\rightarrow0\), \(\tnl'\rightarrow \lambda/Z\), describing the hydrodynamic-like case with maximal strength of nonlinear interactions; as \(|\sigma_c|\) increases so does \(\tnl'\), indicating a slowing down (and potential weakening) of turbulence; when \(\sigma_c\rightarrow\pm1\), \(\tnl'\rightarrow\infty\), describing the pure Alfv\'en wave propagating without turbulent decay. 

This behavior will have corresponding effects on the turbulence age:
\begin{equation}
    \Atb = \int_{r_1}^{r_2}  \frac{dr} {U(r)\tnl'(r)}  \equiv \int_{r_1}^{r_2}  \frac{f[\sigma_c(r)]}{U(r)} \frac{Z(r)}{\lambda(r)} dr,
    \label{eq:At2}
\end{equation}
slowing down the aging of turbulence as \(\sigma_c\) increases. We anticipate that this Alfv\'enic modulation will be significant in near-Sun regions and in the high-\(\sigma_c\) polar fast wind. Below, we evaluate and compare the two formulations \(\Ata\) and \(\Atb\) in observational and simulation data.

\section{Results from observations and numerical modeling}\label{sec:Results}
\subsection{Turbulence age evaluated from \textit{ACE} at 1 AU}\label{sec:ace}
We employ \textit{in-situ} observations from \textit{ACE}, situated at \(\text{L}_1\) and spanning the period from 1998-02-05 00:00:00 to 2023-12-31 23:59:00 (UTC). The magnetic field was obtained from the MAG instrument \citep{smith1998ace} and plasma data (ion number density and velocity) from the SWEPAM instrument \citep{mccomas1998solar}. Downloaded magnetic data with 1-s resolution were down-sampled to 1-min cadence, while density and velocity data at 64-s were up-sampled using linear interpolation to match the magnetic field's cadence. Missing data points are represented as `\(NaN\)' and excluded from analysis. The full dataset is partitioned into consecutive non-overlapping 12-hour intervals. 
\(NaNs\) mainly affect density data, which, for our purpose, are only required to compute mean density within an interval (see below). We exclude intervals in which the \(NaN\) fraction in density is above 90\%. In the remaining intervals, the \(NaN\) fraction in the velocity and magnetic field data is below 10\%. The 12-hour interval duration is chosen to include several turbulent correlation times \citep[\(\sim 1\) hour at 1 au;]{bruno2013LRSP}, which is a standard approach when computing turbulence statistics \citep{matthaeus1982JGR}.\footnote{We have repeated our analysis for 6 and 24 hour intervals, and by excluding intervals in which the fraction of \(NaNs\) in density data is above 70\%. The results are not significantly changed in any of these cases.} 

We identify intervals containing interplanetary coronal mass ejections by referencing a catalog compiled by \cite{Richardson2010SoPh}, which has been updated as recently as 2025 \citep{Richardson2024CMElist}; 1,270 such intervals are identified. The remainder of the intervals (10,661 in number) are considered to be representative of the ambient solar wind at 1 AU, which is our focus here. Slow (\(U<400~\kmps\)), medium (\(400 \le U < 600~\kmps\)), and fast (\(U \ge 600~\kmps\)) wind intervals are identified by reference to the mean solar wind speed \(U\) within each interval: \(U\equiv |\bm{U}|\equiv |\langle \tilde{\bm{U}}\rangle|\), where the \(\langle\cdot\rangle\) operator indicates an arithmetic mean computed over an interval and  \(\tilde{\cdot}\) represents a quantity at 1-min cadence (in this case the ion velocity \(\tilde{\bm{U}}\)). This yields 3666, 6020, and 975 intervals for slow, medium, and fast wind, respectively. Magnetic field \(\tilde{\bm{B}}\) is converted to Alfv\'en units: \(\tilde{\bm{B}}_A=\tilde{\bm{B}}/\sqrt{4\pi\rho}\) where \(\rho=m_pn \equiv  m_p\langle\tilde{n}\rangle\) is the mean mass density of the interval, computed from proton mass \(m_p\) and number density \(\tilde{n}\). Hereafter it is understood that the magnetic field is in Alfv\'en units and the subscript `\(A\)' is dropped. 

Magnetic and velocity fluctuations, cross helicity, and the turbulent correlation length are computed using a standard approach \citep{pope2000book,chhiber2021ApJ_psp}. The fluctuations are  \(\tilde{\bm{u}} = \tilde{\bm{U}} - \bm{U}\) and \(\tilde{\bm{b}} = \tilde{\bm{B}} - \bm{B}\), where \(\bm{B}\equiv \langle\tilde{\bm{B}}\rangle\) is the mean magnetic field in an interval. From these, we compute \(\tilde{\bm{z}}_\pm=\tilde{\bm{u}}\pm \tilde{\bm{b}}\), \(Z^2 = \langle\tilde{u}^2\rangle + \langle \tilde{b}^2\rangle\), and \(\sigma_c=(Z_+^2 - Z_-^2)/(Z_+^2 + Z_-^2)\), where \(Z_\pm^2=\langle \tilde{z}_\pm^2\rangle\). The cross helicity is sector rectified so that positive \(\sigma_c\) always indicates outward (anti-Sunward) propagation of Alfv\'en waves \citep{roberts1987JGRb}; i.e., when the mean radial magnetic field for an interval has positive polarity, \(\sigma_c\) for that interval is multiplied by \(-1\). The correlation scale is estimated for each interval by first computing the normalized autocorrelation function \(\mathcal{R}(\tau)=\langle \tilde{\bm{b}}(t) \cdot \tilde{\bm{b}}(t+\tau)\rangle/\langle \tilde{\bm{b}}^2(t) \rangle\) as a function of temporal lag \(\tau\), up to a maximum lag of 6 hours \citep{Blackman1958,matthaeus1982JGR}. The correlation time \(\tau_c\) is estimated as the first lag at which \(\mathcal{R}\) falls below \(1/e\), or the \(e\)-folding time. 
Taylor's ``frozen-in'' hypothesis \citep{taylor1938ProcRSL} is then used to estimate the correlation length: \(\lambda=U\tau_c\).

The above procedure yields a list of \(U,~Z^2,~\lambda,\) and \(\sigma_c\) values at 1 AU.\footnote{Recall that our averaging interval (12 hours) yields information about the energy-containing scales of turbulence, at which the largest turbulent structures (or `eddies') inject energy into the inertial-range cascade \citep{Kiyani2015RSPTA}.} The near-Earth distributions of these quantities have been examined in several previous studies and are often log-normal \citep[e.g.,][]{Bavassano1998JGR,Ruiz2014SoPh,Wang2024ApJ,Chhiber2025arXiv_V-Zcorr}, and the distributions we obtain (not shown) are consistent with these. The mean values for slow, medium, and fast wind may be found in Table \ref{tab:ace}.
%
%

\begin{table}
\centering
\begin{tabular}{ c | c | c | c  } 
  & Slow & Medium & Fast \\ 
  \hhline{-|-|-|-}
 \(U~(\kmps\)) & \(366\) & \(474\) & \(652\) \\ 
 \(Z^2~(\text{km}^2~\text{s}^{-2})\) & \(1560\) & \(3125\) & \(5050\) \\ 
 \(\lambda ~(10^6~\text{km})\) & \(1.15\) & \(1.04\) & \(0.69\) \\ 
 \(\sigma_c\) & \(0.48\) & \(0.58\) & \(0.70\) \\
 \hhline{-|-|-|-}
\end{tabular}
\caption{Mean values in the \textit{ACE} dataset, for slow (\(U<400~\kmps\)), medium (\(400 \le U < 600~\kmps\)), and fast (\(U \ge 600~\kmps\)) wind intervals.}
\label{tab:ace}
\end{table}
%

To compute the \(\At\) we assume that the integrands in Eqs. \eqref{eq:At1} and \eqref{eq:At2} remain constant between \(r_1=0.2\) AU and \(r_2 = 1\) AU.\footnote{We have chosen \(r_1=0.2\) AU for comparison with \textit{PSP} (Sec. \ref{sec:psp}).}
Although a major simplification
, similar assumptions have been widely employed in the solar wind literature to evaluate the collisional age of the solar wind at 1 AU \citep{kasper2008hot,chhiber2016solar}. Moreover, this simplified estimate will be compared with more realistic ones based on \textit{PSP} and modeling below, which may serve to illustrate any deficiency. We may anticipate that since \(U\) and \(\tnl\) are expected to be smaller closer to the Sun \citep{chhiber2016solar,chen2020ApJS,chhiber2021ApJ_psp}, the 1 AU approximation will increase the effective denominator in the integrand, thus reducing the computed \(A_\text{t}\). Conversely, \(f(\sigma_c)\) is also expected to be smaller near the Sun due to large \(|\sigma_c|\) \citep{chhiber2021ApJ_psp}, so using the 1 AU value for this quantity will tend to increase \(\Atb\).

\begin{figure}
    \centering
       \includegraphics[width=.49\textwidth]{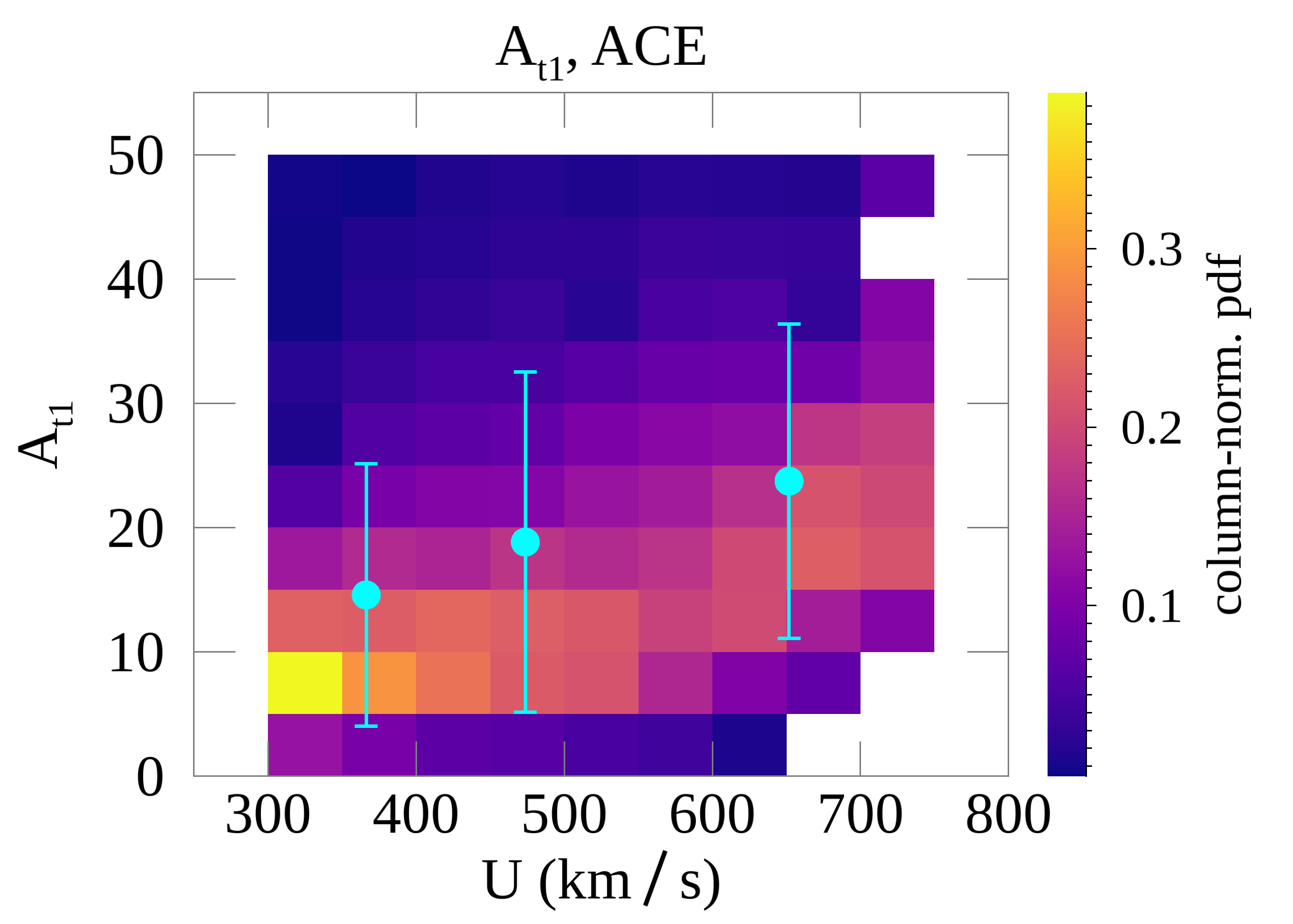}
       \includegraphics[width=.49\textwidth]{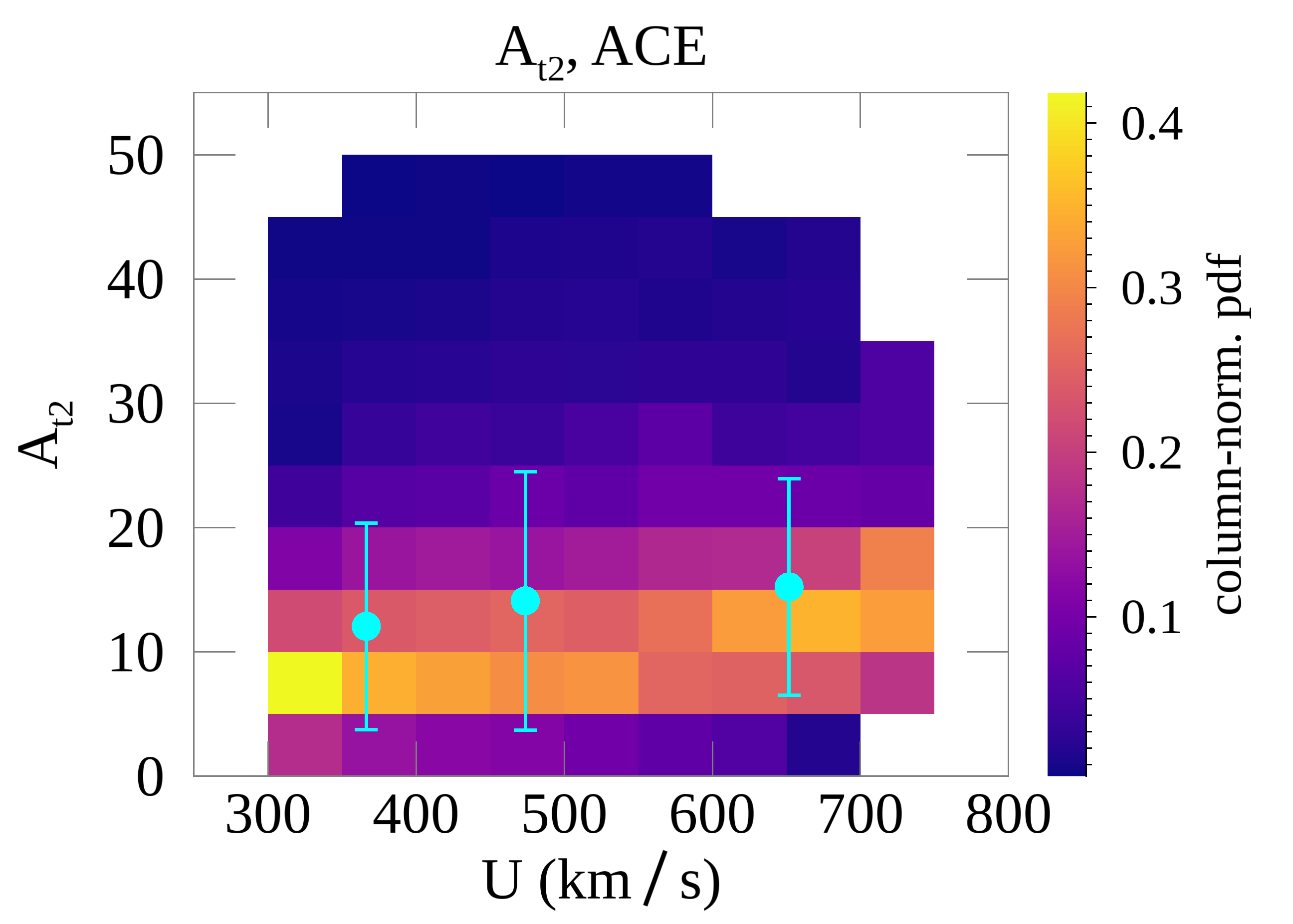}
    \caption{Joint pdfs of turbulence age and solar wind speed, computed from 25 years of \textit{ACE} observations at 1 AU. Bins with fewer than 5 counts are neglected. The pdf of each ``column'', or speed bin, has been normalized by the total counts in that bin. Cyan circles and vertical bars show mean and \(1\sigma\) of \(A_\text{t}\) within slow (\(U< 400 \)), medium (\(400 \le U< 600 \)), and fast (\(U\ge 600\)) intervals. Left and right panels show \(A_\text{t}\) computed using Eqs. \eqref{eq:At1} and \eqref{eq:At2}, respectively, with 1 AU observations (see text). The relative decrease in \(A_\text{t}\) for medium and fast wind when accounting for \(\sigma_c\) (right panel) is evident.}
    \label{fig:ACE_2d_hist}
\end{figure}

We thus evaluate \(\Ata = \Delta r/(U\tnl)=\Delta r~Z/(U\lambda)\) and \(\Atb=\Delta r/(U\tnl')=\Delta r~f(\sigma_c)Z/(U\lambda)\) for each interval, where \(\Delta r=0.8\) AU. Fig. \ref{fig:ACE_2d_hist} shows joint probability distributions (pdfs) of \(U\) and \(A_\text{t1,2}\). The pdf in each speed bin (or `column') has been normalized by the total number of counts in that bin, with colors mapping to the vertical bar that indicates the probability distribution within each speed bin. A clear increasing trend in \(\Ata\) with \(U\) is evident. Cyan symbols show the mean and standard deviation (\(\sigma\)) of \(\At\) in slow, medium, and fast wind: \(\Ata\) is \(\sim 15,~19,\) and \(24\), while \(\Atb\) is \(\sim 12,~14,\) and \(15\) for the three respective speeds, with \(\sigma\) of order 10. Note that for the three wind categories,  \(f(\sigma_c)\) is 0.85, 0.78, and 0.66, respectively. These numbers demonstrate the significant reduction in computed turbulence age when cross-helicity of the Alfv\'enic fast wind is accounted for. Without the modulation factor \(f(\sigma_c)\), the \(\Ata\) calculation would yield the counter-intuitive result that the fast wind is significantly more aged than slow wind \citep[see, e.g.,][]{bruno2013LRSP}. 

Below, we evaluate \(\At\) from \textit{PSP} observations, thus overcoming the ``single-point'' limitation of the \textit{ACE} dataset.
\subsection{Turbulence age evaluated from the \textit{Parker Solar Probe} between 0.2 and 0.8 AU}\label{sec:psp}

We analyze \textit{in-situ} measurements from \textit{PSP}'s orbits 1--25, covering the period from 2018 Oct 31 00:00:00 to 2025 Aug 01 00:00:00 UTC. Although \textit{PSP}'s perihelia extend as low as \(\sim 0.04\) AU, the validity of Taylor's hypothesis becomes questionable below \(\sim 0.2\) AU \citep{chhiber2019psp2,chen2020ApJS}, and we therefore limit our attention to \(r \sim 0.2\) -- 0.8 AU. Magnetic field measurements from FIELDS/MAG \citep{bale2016SSR} and ion moments from SWEAP/SPC \citep{kasper2016SSR,case2020ApJS} instruments are employed. The downloaded MAG data have a time cadence of 3.4 ms, and we interpolate its timeseries to match timestamps of the downloaded ion data. 
The latter's cadence varies through the orbits from \(\sim0.436\) s to 27.5 s, and is relatively higher during solar encounters (\(r\lesssim 0.3\) AU). 
To avoid interpolation of the magnetic field across large data gaps we set it to \(NaN\) if the nearest valid (SPC) timestamp for interpolation is more than 5 s away. SPC data are further filtered by setting values to \(NaN\) at timestamps when data quality flags (\texttt{DQF} or \texttt{general}) are raised.\footnote{See SWEAP Data User Guide at \url{https://cdaweb.gsfc.nasa.gov/pub/data/psp/sweap/sweap_data_user_guide.pdf} for further details.} The full dataset is partitioned into non-overlapping 2-hour intervals.\footnote{A smaller interval duration is chosen compared to \textit{ACE}, to adjust for the shorter correlation times at smaller \(r\) \citep{chen2020ApJS,chhiber2022distrib}.} 
To avoid contamination by missing or low-quality data, we require that at least 60\% of the data in each 2-hour interval have valid non-\(NaN\) ion moments (defined as simultaneous availability of proton density and all three components of bulk velocity). We retain 6615 intervals after these preparations.


In contrast to the 25-year long \textit{ACE} dataset, the \textit{PSP} dataset contains fewer usable intervals, with a much smaller fraction for fast wind. We therefore categorize by just two speed levels: slow (\(U<500~\kmps\)) and fast (\(U\ge500~\kmps\)), yielding 4146 and 398 intervals, respectively. Mean proton speed and density, magnetic field, and turbulence quantities (\(Z^2,~\sigma_c,~\lambda\)) are obtained for each interval as described in Sec. \ref{sec:ace}. A minor difference arises in computation of \(\lambda\): we employ 6-hour intervals instead of the 2-hour ones in order to accommodate a maximum temporal lag of 3 hours, so that the correlation function can be allowed to decorrelate to \(1/e\), and sufficient samples are available for averaging at large lags \citep[e.g.,][]{matthaeus1982JGR}. The three 2-hour intervals constituting the 6-hour interval are each assigned the same value of \(\lambda\).

With these values of \(U,~Z^2,~\lambda,\) and \(\sigma_c\) (at 2-hour cadence) in hand, we bin them in helioradius \(r\) using a bin-width of 0.05 AU. We compute means, \(\sigma\), and quartiles for each quantity, within each radial bin. This procedure allows us to infer radial trends in the quantities of interest, in the sense of a long-term average over \(\sim 7\) years. The radial evolution of the mean values (and \(1\sigma\)) are rather similar to the results shown in \cite{chhiber2021ApJ_psp} based on the first five \textit{PSP} orbits, and are not shown here for brevity. We note here the following overall trends in the radial means observed between 0.2 and 0.8 AU: \(U\) increases from \(\sim 300\) to \(400~\kmps\); \(Z^2\) decreases from \(\sim 10^5\) to \(2\times 10^4~\text{km}^2~\text{s}^{-2}\); \(\lambda\) increases from \(\sim 2\times 10^5\) to \(5\times 10^5\) km; and \(\sigma_c\) decreases from \(\sim0.75\) to \(0.4\).

\begin{figure}
    \centering
    \includegraphics[width=0.49\textwidth]{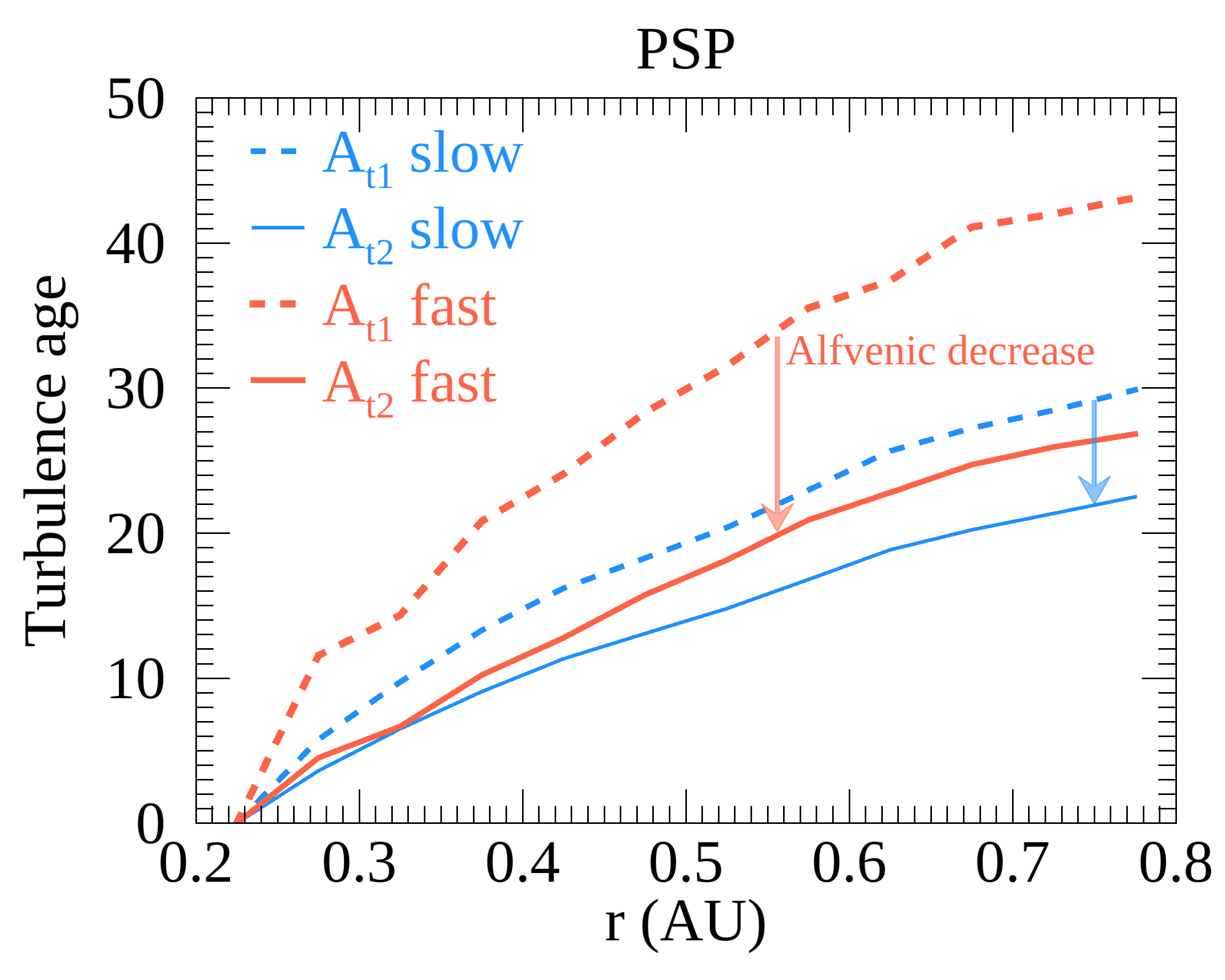}
    \includegraphics[width=0.49\textwidth]{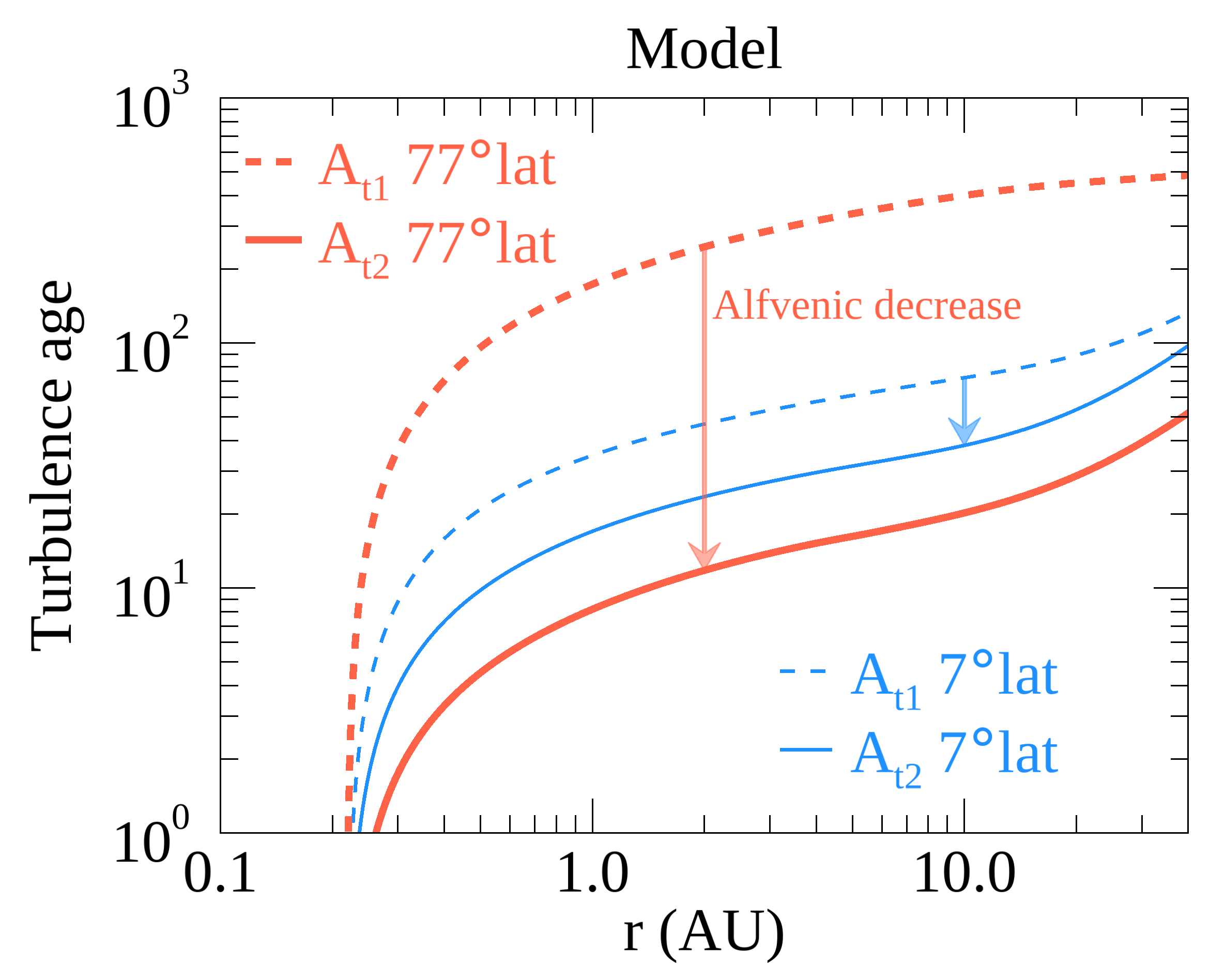}
    \caption{\textit{Left}: Radial evolution of \(A_\text{t}\) in slow and fast solar wind, evaluated from \textit{PSP} as described in the text. \textit{Right}: Radial evolution of \(A_\text{t}\) in simulation, at low (blue) and high (red) heliolatitudes. In both panels \(\Ata\) and \(\Atb\) are evaluated using Eqs. \eqref{eq:At1} and \eqref{eq:At2}, respectively, with the latter accounting for \(\sigma_c\).}
    \label{fig:At_radial}
\end{figure}

To assess the differences between the turbulence age of slow and fast wind, we separately perform the above radial binning for the two categories of intervals.\footnote{Since the solar wind accelerates with \(r\), a potential caveat to note is that the methodology may classify some fast wind (at small \(r\), that has not yet accelerated to a speed above \(500~\kmps\)) as slow. However, this should not be a major effect since most of the acceleration is expected to have already occurred by \(\sim 0.2\) AU \citep[e.g.,][]{Usmanov2025ApJ}.} The most notable distinction (apart from \(U\)) is in \(Z^2\), which is several times larger in the fast wind \citep{Shi2023ApJ,Chhiber2025arXiv_V-Zcorr}. \(\lambda\) and \(\sigma_c\) are both slightly (but systematically) smaller and larger, respectively, in the fast wind. These trends are qualitatively consistent with the ones obtained from \textit{ACE} at 1 AU (Sec. \ref{sec:ace}). We next evaluate Eqs. \eqref{eq:At1} and \eqref{eq:At2}, keeping \(r_1\) fixed at 0.2 AU and increasing \(r_2\) through the mid-points of the radial bins up to 0.8 AU, with the integral evaluated numerically. 
The results are shown in the left panel of Fig. \ref{fig:At_radial}.

We immediately notice the impact that radial variation of solar wind parameters has on the integration: compared to Fig. \ref{fig:ACE_2d_hist}, the computed \(A_\text{t}\) at 0.8 AU is almost twice as large. We also note the decrease in \(A_\text{t}\) when accounting for \(\sigma_c\) (indicated as `Alfv\'enic decrease' in Fig. \ref{fig:At_radial}), similar to the \textit{ACE} calculations. Finally, we note the apparent radial decrease in the rate of increase of \(A_\text{t}\) with \(r\), indicating the slowing of turbulence aging as the solar wind expands through the inner heliosphere. This final point is discussed in more detail in Sec. \ref{sec:pano}. Below, we extend our evaluation of \(\At\) to 40 AU, and beyond the ecliptic, by means of a global solar wind simulation.
\subsection{Turbulence age evaluated from a global solar wind  simulation between 0.2 and 40 AU}\label{sec:glob}
Here we evaluate \(A_\text{t}\) within a three-dimensional MHD model of the global solar wind \citep{Usmanov2025ApJ}. The model couples mean-field equations for bulk velocity, density, magnetic field, and temperature to transport equations for statistical turbulence parameters, including \(Z^2,~\lambda,\) and \(\sigma_c\). This coupling is able to effectively model the heating and acceleration of the solar wind by turbulence \citep{matthaeus2011SSR}, an approach widely employed in global heliospheric modeling \citep{gombosi2018LRSP}. For further details on our model we refer the reader to \cite{Usmanov2025ApJ}, emphasizing here that it is mature, well-tested, and has been validated by comparison with a variety of heliospheric observations \citep[e.g.,][]{usmanov2018,bandyopadhyay2020ApJS_cascade,chhiber2021ApJ_psp,Usmanov2025ApJ}.

For the present study we use a simulation run based on an untilted solar magnetic dipole \citep[described in detail in][]{Usmanov2025ApJ}, rather than photospheric magnetograms corresponding to particular time periods \citep[c.f.][]{chhiber2021ApJ_psp}, to obtain a generic view of typical solar wind conditions (especially during low solar activity). This simulation is symmetric in heliolongitude, and we evaluate \(A_\text{t}\) along two reference radial ``spokes'' at \(\sim 7\degree\) and \(\sim 77\degree\) heliolatitude, corresponding to near-equatorial slow wind and polar fast wind, respectively \citep{mccomas2000JGR,Usmanov2025ApJ}. Eqs. \eqref{eq:At1} and \eqref{eq:At2} are evaluated as described in Sec. \ref{sec:psp}, with \(r_1\) fixed at 0.2 AU and \(r_2\) increased until 40 AU. The results are shown in the right panel of Fig. \ref{fig:At_radial}.

The low-latitude wind has a turbulence age (\(\Ata\) and \(\Atb\)) that behaves quite similarly to the corresponding \textit{PSP} results in the left panel. The polar fast wind, in contrast, undergoes a rapid increase in \(\Ata\) to very large values, reaching above 100 at 1 AU, on account of its high \(Z^2\) and small \(\lambda\) \citep{Usmanov2025ApJ}. On the other hand, accounting for its high \(|\sigma_c|\) with Eq. \eqref{eq:At2} dramatically decreases the computed turbulence age, to values that are around half of the corresponding values for the slow wind. The rate of increase of \(A_\text{t}\) with \(r\) decreases until around 5 AU, beyond which there is an apparent increase in the slope of the curve (see also Fig. \ref{fig:At_radial_pano}, below). We return to this final point below, adding a comparison with \textit{Voyager 1} observations.

\subsection{Radial evolution of turbulence age -- combining \textit{PSP,~ACE,~Voyager 1,} and global simulation results}\label{sec:pano}

\begin{figure}
    \centering
    \includegraphics[width=0.65\textwidth]{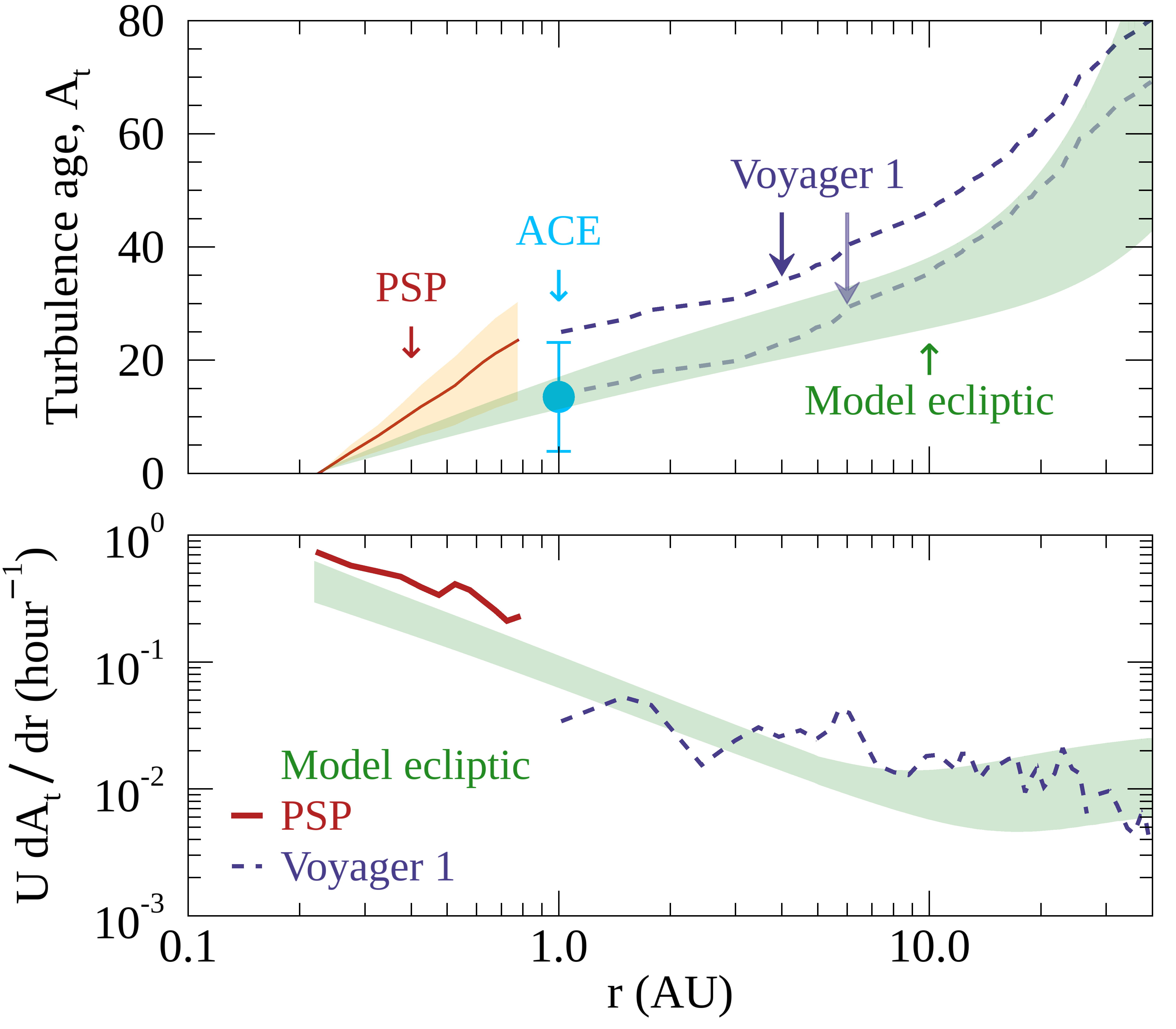}
    \caption{Radial evolution of turbulence age between \(\sim 0.2\) and 40 AU. \textit{PSP}, \textit{ACE}, and global simulation results are computed using Eq. \eqref{eq:At2}, while Voyager 1 result is adapted from Matthaeus et al. 1998. \textit{Top}: Solid dark-red curve shows \(\Atb\) computed from \textit{PSP} observations averaged in radial bins. Tan-shaded region (between 0.2 and 0.8 AU) represents the statistical spread in \(\Atb\) computed from \textit{PSP}. See text for details. Blue circle at 1 AU marks \(\Atb\) computed from \textit{ACE}, with vertical bars marking the \(1\sigma\) spread. Pale-green shaded region between 0.2 and 40 AU shows \(\Atb\) computed from the Usmanov et al. (2025) solar wind model in the ecliptic region, with  upper and lower bounds of the shaded area corresponding to heliolatitudes of \(7\degree\) and \(0\degree\), respectively. Dashed curves show \(A_\text{t}\) adapted from Fig. 2 of Matthaeus et al. (1998); their result has been shifted upwards by ad-hoc amounts (constant in \(r\)) to ``line up'' at 1 AU with the trends from \textit{PSP} and the global simulation (see text).
    \textit{Bottom}: Rate of change of \(A_\text{t}\):  \(\frac{dA_\text{t}}{dt}=U\frac{dA_\text{t}}{dr}\) (see text).}
    \label{fig:At_radial_pano}
\end{figure}

Fig. \ref{fig:At_radial_pano} combines several results from the previous subsections with \textit{Voyager 1} data from M1998, to obtain a panoramic view of the evolution of the turbulence age from near-Sun space to the outer heliosphere. In the top panel, the solid curve shows \(\Atb\) computed from \textit{PSP} observations averaged in radial bins as described in Sec. \ref{sec:psp}, combining slow and fast wind intervals in this instance. The tan-shaded region around this curve represents the statistical spread in the \textit{PSP} data: its upper and lower bounds are obtained by use of the upper and lower quartiles (respectively) of the \(Z^2\) distribution within each radial bin to compute \(\Atb\). The blue symbol indicates the mean and \(1\sigma\) spread of \(\Atb\) computed from \textit{ACE} observations across all wind speeds. The green shaded region represents \(\Atb\) computed from our global solar wind simulation (Sec. \ref{sec:glob}) in the ecliptic region, with  upper and lower bounds of the shaded area corresponding to heliolatitudes of \(7\degree\) and \(0\degree\), respectively. The \textit{Voyager} result (dashed curves) has been adapted from M1998, who computed \(A_\text{t}\) using Eq. \eqref{eq:At1} along the \textit{Voyager 1} trajectory. Naturally, their integration began at \(r_1=1\) AU and their results are therefore agnostic to the accumulated turbulence age from \(r<1\) AU. Accordingly, for our visualization we shift their computed \(A_\text{t}\) upward by (constant in \(r\)) ad-hoc amounts so that the \textit{Voyager} result ``lines up'' at 1 AU with our \textit{PSP} and global simulation results, thus producing the two dashed curves in the top panel of Fig. \ref{fig:At_radial_pano}.\footnote{Even though the global simulation result in Fig. \ref{fig:At_radial_pano} accounts for Alfv\'enicity and the M1998 calculation did not, we have still chosen to present a comparison of the two cases here since the Alfv\'enic modulation is expected to be small in the outer heliosphere, i.e., for most of the \textit{Voyager 1} trajectory, due to relatively small \(\sigma_c\) \citep{Fraternale2022SSR}.} 

The bottom panel of Fig. \ref{fig:At_radial_pano} shows estimates of the rate of change of \(A_\text{t}\) with \(r\): changing variables from time to radial distance (\(dt=dr/U\)), we compute \(dA_\text{t}/dt\) as \(U~ dA_\text{t}/dr\), evaluating the radial derivative numerically for the various curves in the top panel. For the \textit{Voyager} result a constant \(U=400~\kmps\) is assumed.

We observe from Fig. \ref{fig:At_radial_pano} that \(A_\text{t}\) in the global simulation increases at a slower rate compared with the overall trend from \textit{PSP}; while we do not expect a high level of agreement between the generic dipole-based simulation and observations, the slight discrepancy in the two cases is at least partially due to relatively larger (sector rectified) \(\sigma_c\) in the simulation \citep[see Fig. 7 in ][]{chhiber2021ApJ_psp}, which results in greater modulation by the \(f(\sigma_c)\) factor in Eq. \eqref{eq:At2}. The trend from \textit{Voyager} agrees with the model result in general. The bottom panel quantifies the decreasing rate of turbulence aging as the solar wind expands through the inner heliosphere, in both \textit{PSP} and simulation data; the \textit{Voyager} curve also indicates a decrease (albeit with fluctuations, although these may be amplified by the logarithmic \(r\)-axis). Returning to the top panel, the \textit{Voyager} result apparently agrees quite well with the lower bound from the simulation between 1 and 4 AU. Around 5 AU, both \textit{Voyager} and simulation curves become flatter, with the simulation indicating a clear increase in \(dA_\text{t}/dt\) as the solar wind expands into the outer heliosphere. This trend, which was also discussed by M1998, is likely due to \textit{in-situ} driving of solar wind turbulence by pickup ions, an effect which is modeled in our simulations \citep[see][]{Usmanov2025ApJ}.

\section{Discussion and Conclusions}\label{sec:disc}

In this paper we have proposed a new formulation for estimating the turbulence age of the solar wind that accounts for the varying levels of Alfv\'enicity (specifically, cross helicity) that are routinely observed in the heliosphere. Our work builds on \cite{matthaeus1998JGR}, who introduced this type of quantitative determination of turbulence age to space physics studies, but left an accounting of non-zero \(\sigma_c\) to future work. We evaluated \(A_\text{t}\) based on the two formulations, employing data from \textit{PSP}, \textit{ACE}, \textit{Voyager 1}, and a global solar wind simulation, spanning distances of 0.2 to 40 AU from the Sun, and including both ecliptic and polar  wind.

We find that neglect of \(\sigma_c\) may produce unreasonably large estimates of \(A_\text{t}\) for the fast solar wind (especially in polar regions); the new formulation yields values indicating a similar level of turbulence development across wind speeds, which is a noteworthy result in itself. In the ecliptic (\textit{ACE} and \textit{PSP}), the fast wind is slightly more aged than the slow wind, while the polar wind (evaluated in the global simulation) has a smaller turbulence age than equatorial wind. These findings add nuance to the view that slow wind is generically ``more turbulent'' than fast wind \citep[e.g.,][]{dasso2005ApJ,Wang2024ApJ}. They may also provide insight on the role of turbulence in heating and accelerating the fast wind, in view of the observed positive correlations between speed, proton temperature, and fluctuation amplitude  \citep{Demoulin2009SoPh,Elliott2012JGR,Shi2023ApJ,Chhiber2025arXiv_V-Zcorr}.

We also examined trends in the evolution of turbulence age as the solar wind expands through the inner and outer heliospheres. Around 20 nonlinear times may pass between 0.2 and 1 AU, indicating well developed turbulence, although the rate of development is much larger near the Sun. Around 5 AU, both \textit{Voyager} and global simulation results show an increase in the rate of aging, likely related to \textit{in-situ} driving of turbulence by pickup ions \citep{Fraternale2022SSR}. Our analysis also suggests that the state of the turbulence at 1 AU is both \textit{source} and \textit{path} dependent: slow and fast wind \citep[originating in different solar sources; e.g.,][]{Schwenn2006SSR_wind_sources} may have different \(A_\text{t}\), but neglecting the variation of various parameters during the passage of the solar wind may lead to underestimation of the turbulence age (compare \(\Atb\) from \textit{ACE} and \textit{PSP} in Sec. \ref{sec:Results}).

These results may help interpret observations of other turbulence parameters and their complex radial evolution. E.g., the Reynolds number appears to decrease between \(\sim 0.1\) to 10 AU \citep{Cuesta2022ApJS,Adhikari2025ApJ}, while small-scale intermittency increases in the young solar wind \citep[\(r<1\) AU;][]{alberti2020psp,Telloni2021ApJ} before decreasing between \(\sim 1\) and 10 AU \citep{Cuesta2022ApJS}. Further signatures of turbulence aging with distance are found in the steepening inertial-range spectral slope and increase in correlation scale \citep{chen2020ApJS,Cuesta2022ApJL}, along with reduction in cross helicity \citep{roberts1992jgr}.
Better understanding 
of times scales and aging of turbulence in the outer heliosphere 
may be achieved by taking into account the developments reported above. This may be
particularly relevant to the recently launched \textit{IMAP} mission
\citep{MccomasEA25-IMAP}
which seeks to understand
the influence of 1 AU conditions on the outer heliospheric boundaries and related energetic particles.

In future work, it may be of interest to revisit interpretations that were based on the definition of turbulence age that neglected cross helicity \citep[e.g.,][]{McIntyre2023ApJ}. The turbulence age can also be evaluated from observations outside the ecliptic from \textit{Ulysses} and \textit{Solar Orbiter} \citep{mccomas2000JGR,Muller2020AA}, and compared with the type of global simulations we analyzed here. The turbulence age in the sub-Alfv\'enic corona observed by \textit{PSP} would also be of great interest \citep[e.g.,][]{bandyopadhyay2022ApJ}, although the interpretation may be complicated by issues relating to Taylor's frozen-in hypothesis and the sampling direction bias at \textit{PSP}'s perihelia \citep{chhiber2019psp2,Cuesta2022ApJL}.


\section*{Acknowledgments}
This research was supported by NASA under the Living With a Star (LWS) Science program grant 80NSSC22K1020, the Heliophysics Supporting Research (HSR) program grant 80NSSC22K1639, and the \textit{IMAP} mission through a subcontract from Princeton University (SUB0000317). It utilized resources provided by the  National Energy Research Scientific Computing Center. \textit{ACE} magnetic field and plasma data were downloaded
from \url{https://spdf.gsfc.nasa.gov/pub/data/ace/mag/level_2_cdaweb/mfi_h3/} and \url{https://spdf.gsfc.nasa.gov/pub/data/ace/swepam/level2_hdf/ions_64sec}, respectively. The ICME list is available at \url{https://doi.org/10.7910/DVN/C2MHTH}. \textit{PSP} data (level-2 magnetic-field and level-3 ion-moment data products) were downloaded from the Heliodata website (\url{https://helio.data.nasa.gov/mission/ParkerSolarProbe}).


\begin{thebibliography}{61}
\providecommand{\natexlab}[1]{#1}
\providecommand{\url}[1]{\texttt{#1}}
\expandafter\ifx\csname urlstyle\endcsname\relax
  \providecommand{\doi}[1]{doi: #1}\else
  \providecommand{\doi}{doi: \begingroup \urlstyle{rm}\Url}\fi

\bibitem[{Adhikari} et~al.(2025){Adhikari}, {Giri}, {Karki}, {Baruwal}, {Baruwal}, {Khondoker Shikha}, {Mainali}, {Kenno}, {Tasnim}, {Zank}, {Wang}, {Treville}, and {Ghimire}]{Adhikari2025ApJ}
Laxman {Adhikari}, Ashutosh {Giri}, Monika {Karki}, Prashant {Baruwal}, Prashrit {Baruwal}, Rubaiya {Khondoker Shikha}, Bigyan {Mainali}, Dessalegn {Kenno}, Ismita {Tasnim}, Gary~P. {Zank}, Bingbing {Wang}, Lucien {Treville}, and Sagar {Ghimire}.
\newblock {Reynolds Number and Viscosity in the Solar Wind}.
\newblock \emph{\apj}, 993\penalty0 (1):\penalty0 60, November 2025.
\newblock \doi{10.3847/1538-4357/ae101d}.

\bibitem[Alberti et~al.(2020)Alberti, Laurenza, Consolini, Milillo, Marcucci, Carbone, and Bale]{alberti2020psp}
Tommaso Alberti, Monica Laurenza, Giuseppe Consolini, Anna Milillo, Maria~Federica Marcucci, Vincenzo Carbone, and Stuart~D. Bale.
\newblock On the scaling properties of magnetic-field fluctuations through the inner heliosphere.
\newblock \emph{The Astrophysical Journal}, 902\penalty0 (1):\penalty0 84, oct 2020.
\newblock \doi{10.3847/1538-4357/abb3d2}.
\newblock URL \url{https://doi.org/10.3847/1538-4357/abb3d2}.

\bibitem[{Bale} et~al.(2016){Bale}, {Goetz}, {Harvey}, {Turin}, {Bonnell}, {Dudok de Wit}, {Ergun}, {MacDowall}, {Pulupa}, {Andre}, {Bolton}, {Bougeret}, {Bowen}, {Burgess}, {Cattell}, {Chandran}, {Chaston}, {Chen}, {Choi}, {Connerney}, {Cranmer}, {Diaz-Aguado}, {Donakowski}, {Drake}, {Farrell}, {Fergeau}, {Fermin}, {Fischer}, {Fox}, {Glaser}, {Goldstein}, {Gordon}, {Hanson}, {Harris}, {Hayes}, {Hinze}, {Hollweg}, {Horbury}, {Howard}, {Hoxie}, {Jannet}, {Karlsson}, {Kasper}, {Kellogg}, {Kien}, {Klimchuk}, {Krasnoselskikh}, {Krucker}, {Lynch}, {Maksimovic}, {Malaspina}, {Marker}, {Martin}, {Martinez-Oliveros}, {McCauley}, {McComas}, {McDonald}, {Meyer-Vernet}, {Moncuquet}, {Monson}, {Mozer}, {Murphy}, {Odom}, {Oliverson}, {Olson}, {Parker}, {Pankow}, {Phan}, {Quataert}, {Quinn}, {Ruplin}, {Salem}, {Seitz}, {Sheppard}, {Siy}, {Stevens}, {Summers}, {Szabo}, {Timofeeva}, {Vaivads}, {Velli}, {Yehle}, {Werthimer}, and {Wygant}]{bale2016SSR}
S.~D. {Bale}, K.~{Goetz}, P.~R. {Harvey}, P.~{Turin}, J.~W. {Bonnell}, T.~{Dudok de Wit}, R.~E. {Ergun}, R.~J. {MacDowall}, M.~{Pulupa}, M.~{Andre}, M.~{Bolton}, J.-L. {Bougeret}, T.~A. {Bowen}, D.~{Burgess}, C.~A. {Cattell}, B.~D.~G. {Chandran}, C.~C. {Chaston}, C.~H.~K. {Chen}, M.~K. {Choi}, J.~E. {Connerney}, S.~{Cranmer}, M.~{Diaz-Aguado}, W.~{Donakowski}, J.~F. {Drake}, W.~M. {Farrell}, P.~{Fergeau}, J.~{Fermin}, J.~{Fischer}, N.~{Fox}, D.~{Glaser}, M.~{Goldstein}, D.~{Gordon}, E.~{Hanson}, S.~E. {Harris}, L.~M. {Hayes}, J.~J. {Hinze}, J.~V. {Hollweg}, T.~S. {Horbury}, R.~A. {Howard}, V.~{Hoxie}, G.~{Jannet}, M.~{Karlsson}, J.~C. {Kasper}, P.~J. {Kellogg}, M.~{Kien}, J.~A. {Klimchuk}, V.~V. {Krasnoselskikh}, S.~{Krucker}, J.~J. {Lynch}, M.~{Maksimovic}, D.~M. {Malaspina}, S.~{Marker}, P.~{Martin}, J.~{Martinez-Oliveros}, J.~{McCauley}, D.~J. {McComas}, T.~{McDonald}, N.~{Meyer-Vernet}, M.~{Moncuquet}, S.~J. {Monson}, F.~S. {Mozer}, S.~D. {Murphy}, J.~{Odom}, R.~{Oliverson}, J.~{Olson}, E.~N. {Parker},
  D.~{Pankow}, T.~{Phan}, E.~{Quataert}, T.~{Quinn}, S.~W. {Ruplin}, C.~{Salem}, D.~{Seitz}, D.~A. {Sheppard}, A.~{Siy}, K.~{Stevens}, D.~{Summers}, A.~{Szabo}, M.~{Timofeeva}, A.~{Vaivads}, M.~{Velli}, A.~{Yehle}, D.~{Werthimer}, and J.~R. {Wygant}.
\newblock {The FIELDS Instrument Suite for Solar Probe Plus. Measuring the Coronal Plasma and Magnetic Field, Plasma Waves and Turbulence, and Radio Signatures of Solar Transients}.
\newblock \emph{\ssr}, 204:\penalty0 49--82, December 2016.
\newblock \doi{10.1007/s11214-016-0244-5}.

\bibitem[{Bandyopadhyay} et~al.(2022){Bandyopadhyay}, {Matthaeus}, {McComas}, {Chhiber}, {Usmanov}, {Huang}, {Livi}, {Larson}, {Kasper}, {Case}, {Stevens}, {Whittlesey}, {Romeo}, {Bale}, {Bonnell}, {Dudok de Wit}, {Goetz}, {Harvey}, {MacDowall}, {Malaspina}, and {Pulupa}]{bandyopadhyay2022ApJ}
R.~{Bandyopadhyay}, W.~H. {Matthaeus}, D.~J. {McComas}, R.~{Chhiber}, A.~V. {Usmanov}, J.~{Huang}, R.~{Livi}, D.~E. {Larson}, J.~C. {Kasper}, A.~W. {Case}, M.~{Stevens}, P.~{Whittlesey}, O.~M. {Romeo}, S.~D. {Bale}, J.~W. {Bonnell}, T.~{Dudok de Wit}, K.~{Goetz}, P.~R. {Harvey}, R.~J. {MacDowall}, D.~M. {Malaspina}, and M.~{Pulupa}.
\newblock {Sub-Alfv{\'e}nic Solar Wind Observed by the Parker Solar Probe: Characterization of Turbulence, Anisotropy, Intermittency, and Switchback}.
\newblock \emph{\apjl}, 926\penalty0 (1):\penalty0 L1, February 2022.
\newblock \doi{10.3847/2041-8213/ac4a5c}.

\bibitem[{Bandyopadhyay} et~al.(2020){Bandyopadhyay}, {Goldstein}, {Maruca}, {Matthaeus}, {Parashar}, {Ruffolo}, {Chhiber}, {Usmanov}, {Chasapis}, {Qudsi}, {Bale}, {Bonnell}, {Dudok de Wit}, {Goetz}, {Harvey}, {MacDowall}, {Malaspina}, {Pulupa}, {Kasper}, {Korreck}, {Case}, {Stevens}, {Whittlesey}, {Larson}, {Livi}, {Klein}, {Velli}, and {Raouafi}]{bandyopadhyay2020ApJS_cascade}
Riddhi {Bandyopadhyay}, M.~L. {Goldstein}, B.~A. {Maruca}, W.~H. {Matthaeus}, T.~N. {Parashar}, D.~{Ruffolo}, R.~{Chhiber}, A.~{Usmanov}, A.~{Chasapis}, R.~{Qudsi}, Stuart~D. {Bale}, J.~W. {Bonnell}, Thierry {Dudok de Wit}, Keith {Goetz}, Peter~R. {Harvey}, Robert~J. {MacDowall}, David~M. {Malaspina}, Marc {Pulupa}, J.~C. {Kasper}, K.~E. {Korreck}, A.~W. {Case}, M.~{Stevens}, P.~{Whittlesey}, D.~{Larson}, R.~{Livi}, K.~G. {Klein}, M.~{Velli}, and N.~{Raouafi}.
\newblock {Enhanced Energy Transfer Rate in Solar Wind Turbulence Observed near the Sun from Parker Solar Probe}.
\newblock \emph{\apjs}, 246\penalty0 (2):\penalty0 48, February 2020.
\newblock \doi{10.3847/1538-4365/ab5dae}.

\bibitem[{Batchelor}(1953)]{Batchelor1953book}
G.~K. {Batchelor}.
\newblock \emph{{The Theory of Homogeneous Turbulence}}.
\newblock Cambridge University Press, 1953.

\bibitem[{Bavassano} et~al.(1998){Bavassano}, {Pietropaolo}, and {Bruno}]{Bavassano1998JGR}
B.~{Bavassano}, E.~{Pietropaolo}, and R.~{Bruno}.
\newblock {Cross-helicity and residual energy in solar wind turbulence: Radial evolution and latitudinal dependence in the region from 1 to 5 AU}.
\newblock \emph{\jgr}, 103\penalty0 (A4):\penalty0 6521--6530, April 1998.
\newblock \doi{10.1029/97JA03029}.

\bibitem[{Belcher} and {Davis}(1971)]{belcher1971JGR}
J.~W. {Belcher} and L.~{Davis}, Jr.
\newblock {Large-amplitude Alfv{\'e}n waves in the interplanetary medium, 2}.
\newblock \emph{\jgr}, 76:\penalty0 3534, 1971.
\newblock \doi{10.1029/JA076i016p03534}.

\bibitem[Blackman and Tukey(1958)]{Blackman1958}
R.~B. Blackman and J.~W. Tukey.
\newblock The measurement of power spectra from the point of view of communications engineering — part i.
\newblock \emph{The Bell System Technical Journal}, 37\penalty0 (1):\penalty0 185--282, 1958.
\newblock \doi{10.1002/j.1538-7305.1958.tb03874.x}.

\bibitem[{Breech} et~al.(2008){Breech}, {Matthaeus}, {Minnie}, {Bieber}, {Oughton}, {Smith}, and {Isenberg}]{breech2008turbulence}
B.~{Breech}, W.~H. {Matthaeus}, J.~{Minnie}, J.~W. {Bieber}, S.~{Oughton}, C.~W. {Smith}, and P.~A. {Isenberg}.
\newblock {Turbulence transport throughout the heliosphere}.
\newblock \emph{Journal of Geophysical Research (Space Physics)}, 113:\penalty0 A08105, August 2008.
\newblock \doi{10.1029/2007JA012711}.

\bibitem[{Bruno} and {Carbone}(2013)]{bruno2013LRSP}
R.~{Bruno} and V.~{Carbone}.
\newblock {The Solar Wind as a Turbulence Laboratory}.
\newblock \emph{Living Reviews in Solar Physics}, 10:\penalty0 2, December 2013.
\newblock \doi{10.12942/lrsp-2013-2}.

\bibitem[{Case} et~al.(2020){Case}, {Kasper}, {Stevens}, {Korreck}, {Paulson}, {Daigneau}, {Caldwell}, {Freeman}, {Henry}, {Klingensmith}, {Bookbinder}, {Robinson}, {Berg}, {Tiu}, {Wright}, {Reinhart}, {Curtis}, {Ludlam}, {Larson}, {Whittlesey}, {Livi}, {Klein}, and {Martinovi{\'c}}]{case2020ApJS}
A.~W. {Case}, Justin~C. {Kasper}, Michael~L. {Stevens}, Kelly~E. {Korreck}, Kristoff {Paulson}, Peter {Daigneau}, Dave {Caldwell}, Mark {Freeman}, Thayne {Henry}, Brianna {Klingensmith}, J.~A. {Bookbinder}, Miles {Robinson}, Peter {Berg}, Chris {Tiu}, Jr. {Wright}, K.~H., Matthew~J. {Reinhart}, David {Curtis}, Michael {Ludlam}, Davin {Larson}, Phyllis {Whittlesey}, Roberto {Livi}, Kristopher~G. {Klein}, and Mihailo~M. {Martinovi{\'c}}.
\newblock {The Solar Probe Cup on the Parker Solar Probe}.
\newblock \emph{\apjs}, 246\penalty0 (2):\penalty0 43, February 2020.
\newblock \doi{10.3847/1538-4365/ab5a7b}.

\bibitem[{Chen} et~al.(2020){Chen}, {Bale}, {Bonnell}, {Borovikov}, {Bowen}, {Burgess}, {Case}, {Chandran}, {de Wit}, {Goetz}, {Harvey}, {Kasper}, {Klein}, {Korreck}, {Larson}, {Livi}, {MacDowall}, {Malaspina}, {Mallet}, {McManus}, {Moncuquet}, {Pulupa}, {Stevens}, and {Whittlesey}]{chen2020ApJS}
C.~H.~K. {Chen}, S.~D. {Bale}, J.~W. {Bonnell}, D.~{Borovikov}, T.~A. {Bowen}, D.~{Burgess}, A.~W. {Case}, B.~D.~G. {Chandran}, T.~Dudok {de Wit}, K.~{Goetz}, P.~R. {Harvey}, J.~C. {Kasper}, K.~G. {Klein}, K.~E. {Korreck}, D.~{Larson}, R.~{Livi}, R.~J. {MacDowall}, D.~M. {Malaspina}, A.~{Mallet}, M.~D. {McManus}, M.~{Moncuquet}, M.~{Pulupa}, M.~L. {Stevens}, and P.~{Whittlesey}.
\newblock {The Evolution and Role of Solar Wind Turbulence in the Inner Heliosphere}.
\newblock \emph{\apjs}, 246\penalty0 (2):\penalty0 53, February 2020.
\newblock \doi{10.3847/1538-4365/ab60a3}.

\bibitem[{Chhiber} et~al.(2016){Chhiber}, {Usmanov}, {Matthaeus}, and {Goldstein}]{chhiber2016solar}
R.~{Chhiber}, A.~{Usmanov}, W.~{Matthaeus}, and M.~{Goldstein}.
\newblock {Solar Wind Collisional Age from a Global Magnetohydrodynamics Simulation}.
\newblock \emph{\apj}, 821:\penalty0 34, April 2016.
\newblock \doi{10.3847/0004-637X/821/1/34}.

\bibitem[{Chhiber}(2022)]{chhiber2022distrib}
Rohit {Chhiber}.
\newblock {Distributions of Magnetic Field Components in the Inner Solar Wind -- Parker Solar Probe Observations}.
\newblock \emph{In Prep.}, 2022.

\bibitem[{Chhiber} et~al.(2019){Chhiber}, {Usmanov}, {Matthaeus}, {Parashar}, and {Goldstein}]{chhiber2019psp2}
Rohit {Chhiber}, Arcadi~V. {Usmanov}, William~H. {Matthaeus}, Tulasi~N. {Parashar}, and Melvyn~L. {Goldstein}.
\newblock {Contextual Predictions for Parker Solar Probe. II. Turbulence Properties and Taylor Hypothesis}.
\newblock \emph{\apjs}, 242\penalty0 (1):\penalty0 12, May 2019.
\newblock \doi{10.3847/1538-4365/ab16d7}.

\bibitem[{Chhiber} et~al.(2021){Chhiber}, {Usmanov}, {Matthaeus}, and {Goldstein}]{chhiber2021ApJ_psp}
Rohit {Chhiber}, Arcadi~V. {Usmanov}, William~H. {Matthaeus}, and Melvyn~L. {Goldstein}.
\newblock {Large-scale Structure and Turbulence Transport in the Inner Solar Wind: Comparison of Parker Solar Probe's First Five Orbits with a Global 3D Reynolds-averaged MHD Model}.
\newblock \emph{\apj}, 923\penalty0 (1):\penalty0 89, December 2021.
\newblock \doi{10.3847/1538-4357/ac1ac7}.

\bibitem[{Chhiber} et~al.(2025){Chhiber}, {Wang}, {Wang}, and {Roy}]{Chhiber2025arXiv_V-Zcorr}
Rohit {Chhiber}, Yanwen {Wang}, Jiaming {Wang}, and Sohom {Roy}.
\newblock {A robust empirical relationship between speed and turbulence energy in the near-Earth solar wind}.
\newblock \emph{arXiv e-prints}, art. arXiv:2512.00983, November 2025.
\newblock \doi{10.48550/arXiv.2512.00983}.

\bibitem[{Cuesta} et~al.(2022{\natexlab{a}}){Cuesta}, {Chhiber}, {Roy}, {Goodwill}, {Pecora}, {Jarosik}, {Matthaeus}, {Parashar}, and {Bandyopadhyay}]{Cuesta2022ApJL}
Manuel~Enrique {Cuesta}, Rohit {Chhiber}, Sohom {Roy}, Joshua {Goodwill}, Francesco {Pecora}, Jake {Jarosik}, William~H. {Matthaeus}, Tulasi~N. {Parashar}, and Riddhi {Bandyopadhyay}.
\newblock {Isotropization and Evolution of Energy-containing Eddies in Solar Wind Turbulence: Parker Solar Probe, Helios 1, ACE, WIND, and Voyager 1}.
\newblock \emph{\apjl}, 932\penalty0 (1):\penalty0 L11, June 2022{\natexlab{a}}.
\newblock \doi{10.3847/2041-8213/ac73fd}.

\bibitem[{Cuesta} et~al.(2022{\natexlab{b}}){Cuesta}, {Parashar}, {Chhiber}, and {Matthaeus}]{Cuesta2022ApJS}
Manuel~Enrique {Cuesta}, Tulasi~N. {Parashar}, Rohit {Chhiber}, and William~H. {Matthaeus}.
\newblock {Intermittency in the Expanding Solar Wind: Observations from Parker Solar Probe (0.16 au), Helios 1 (0.3-1 au), and Voyager 1 (1-10 au)}.
\newblock \emph{\apjs}, 259\penalty0 (1):\penalty0 23, March 2022{\natexlab{b}}.
\newblock \doi{10.3847/1538-4365/ac45fa}.

\bibitem[{Dasso} et~al.(2005){Dasso}, {Milano}, {Matthaeus}, and {Smith}]{dasso2005ApJ}
S.~{Dasso}, L.~J. {Milano}, W.~H. {Matthaeus}, and C.~W. {Smith}.
\newblock {Anisotropy in Fast and Slow Solar Wind Fluctuations}.
\newblock \emph{\apjl}, 635\penalty0 (2):\penalty0 L181--L184, December 2005.
\newblock \doi{10.1086/499559}.

\bibitem[{de K\'arm\'an} and {Howarth}(1938)]{karman1938prsl}
T.~{de K\'arm\'an} and L.~{Howarth}.
\newblock {On the Statistical Theory of Isotropic Turbulence}.
\newblock \emph{Proceedings of the Royal Society of London Series A}, 164:\penalty0 192--215, January 1938.
\newblock \doi{10.1098/rspa.1938.0013}.

\bibitem[{De Pontieu} et~al.(2007){De Pontieu}, {McIntosh}, {Carlsson}, {Hansteen}, {Tarbell}, {Schrijver}, {Title}, {Shine}, {Tsuneta}, {Katsukawa}, {Ichimoto}, {Suematsu}, {Shimizu}, and {Nagata}]{DePontieu2007Sci}
B.~{De Pontieu}, S.~W. {McIntosh}, M.~{Carlsson}, V.~H. {Hansteen}, T.~D. {Tarbell}, C.~J. {Schrijver}, A.~M. {Title}, R.~A. {Shine}, S.~{Tsuneta}, Y.~{Katsukawa}, K.~{Ichimoto}, Y.~{Suematsu}, T.~{Shimizu}, and S.~{Nagata}.
\newblock {Chromospheric Alfv{\'e}nic Waves Strong Enough to Power the Solar Wind}.
\newblock \emph{Science}, 318\penalty0 (5856):\penalty0 1574, December 2007.
\newblock \doi{10.1126/science.1151747}.

\bibitem[{D{\'e}moulin}(2009)]{Demoulin2009SoPh}
P.~{D{\'e}moulin}.
\newblock {Why Do Temperature and Velocity Have Different Relationships in the Solar Wind and in Interplanetary Coronal Mass Ejections?}
\newblock \emph{\solphys}, 257\penalty0 (1):\penalty0 169--184, June 2009.
\newblock \doi{10.1007/s11207-009-9338-5}.

\bibitem[{Dobrowolny} et~al.(1980){Dobrowolny}, {Mangeney}, and {Veltri}]{Dobrowolny1980PRL}
M.~{Dobrowolny}, A.~{Mangeney}, and P.~{Veltri}.
\newblock {Fully Developed Anisotropic Hydromagnetic Turbulence in Interplanetary Space}.
\newblock \emph{\prl}, 45\penalty0 (2):\penalty0 144--147, July 1980.
\newblock \doi{10.1103/PhysRevLett.45.144}.

\bibitem[{Elliott} et~al.(2012){Elliott}, {Henney}, {McComas}, {Smith}, and {Vasquez}]{Elliott2012JGR}
H.~A. {Elliott}, C.~J. {Henney}, D.~J. {McComas}, C.~W. {Smith}, and B.~J. {Vasquez}.
\newblock {Temporal and radial variation of the solar wind temperature-speed relationship}.
\newblock \emph{Journal of Geophysical Research (Space Physics)}, 117\penalty0 (A9):\penalty0 A09102, September 2012.
\newblock \doi{10.1029/2011JA017125}.

\bibitem[{Els\"asser}(1950)]{elsasser1950PhRv}
W.~M. {Els\"asser}.
\newblock {The Hydromagnetic Equations}.
\newblock \emph{Physical Review}, 79:\penalty0 183--183, July 1950.
\newblock \doi{10.1103/PhysRev.79.183}.

\bibitem[{Engelbrecht} et~al.(2022){Engelbrecht}, {Effenberger}, {Florinski}, {Potgieter}, {Ruffolo}, {Chhiber}, {Usmanov}, {Rankin}, and {Els}]{engelbrecht2022ssr}
N.~Eugene {Engelbrecht}, F.~{Effenberger}, V.~{Florinski}, M.~S. {Potgieter}, D.~{Ruffolo}, R.~{Chhiber}, A.~V. {Usmanov}, J.~S. {Rankin}, and P.~L. {Els}.
\newblock {Theory of Cosmic Ray Transport in the Heliosphere}.
\newblock \emph{\ssr}, 218\penalty0 (4):\penalty0 33, June 2022.
\newblock \doi{10.1007/s11214-022-00896-1}.

\bibitem[{Fraternale} et~al.(2022){Fraternale}, {Adhikari}, {Fichtner}, {Kim}, {Kleimann}, {Oughton}, {Pogorelov}, {Roytershteyn}, {Smith}, {Usmanov}, {Zank}, and {Zhao}]{Fraternale2022SSR}
Federico {Fraternale}, Laxman {Adhikari}, Horst {Fichtner}, Tae~K. {Kim}, Jens {Kleimann}, Sean {Oughton}, Nikolai~V. {Pogorelov}, Vadim {Roytershteyn}, Charles~W. {Smith}, Arcadi~V. {Usmanov}, Gary~P. {Zank}, and Lingling {Zhao}.
\newblock {Turbulence in the Outer Heliosphere}.
\newblock \emph{\ssr}, 218\penalty0 (6):\penalty0 50, September 2022.
\newblock \doi{10.1007/s11214-022-00914-2}.

\bibitem[{Gombosi} et~al.(2018){Gombosi}, {van der Holst}, {Manchester}, and {Sokolov}]{gombosi2018LRSP}
T.~I. {Gombosi}, B.~{van der Holst}, W.~B. {Manchester}, and I.~V. {Sokolov}.
\newblock {Extended MHD modeling of the steady solar corona and the solar wind}.
\newblock \emph{Living Reviews in Solar Physics}, 15:\penalty0 4, July 2018.
\newblock \doi{10.1007/s41116-018-0014-4}.

\bibitem[{Hossain} et~al.(1995){Hossain}, {Gray}, {Pontius}, {Matthaeus}, and {Oughton}]{hossain1995PhFl}
M.~{Hossain}, P.~C. {Gray}, D.~H. {Pontius}, Jr., W.~H. {Matthaeus}, and S.~{Oughton}.
\newblock {Phenomenology for the decay of energy-containing eddies in homogeneous MHD turbulence}.
\newblock \emph{Physics of Fluids}, 7:\penalty0 2886--2904, November 1995.
\newblock \doi{10.1063/1.868665}.

\bibitem[{Kasper} et~al.(2008){Kasper}, {Lazarus}, and {Gary}]{kasper2008hot}
J.~C. {Kasper}, A.~J. {Lazarus}, and S.~P. {Gary}.
\newblock {Hot Solar-Wind Helium: Direct Evidence for Local Heating by Alfv{\'e}n-Cyclotron Dissipation}.
\newblock \emph{Physical Review Letters}, 101\penalty0 (26):\penalty0 261103, December 2008.
\newblock \doi{10.1103/PhysRevLett.101.261103}.

\bibitem[{Kasper} et~al.(2016){Kasper}, {Abiad}, {Austin}, {Balat-Pichelin}, {Bale}, {Belcher}, {Berg}, {Bergner}, {Berthomier}, {Bookbinder}, {Brodu}, {Caldwell}, {Case}, {Chandran}, {Cheimets}, {Cirtain}, {Cranmer}, {Curtis}, {Daigneau}, {Dalton}, {Dasgupta}, {DeTomaso}, {Diaz-Aguado}, {Djordjevic}, {Donaskowski}, {Effinger}, {Florinski}, {Fox}, {Freeman}, {Gallagher}, {Gary}, {Gauron}, {Gates}, {Goldstein}, {Golub}, {Gordon}, {Gurnee}, {Guth}, {Halekas}, {Hatch}, {Heerikuisen}, {Ho}, {Hu}, {Johnson}, {Jordan}, {Korreck}, {Larson}, {Lazarus}, {Li}, {Livi}, {Ludlam}, {Maksimovic}, {McFadden}, {Marchant}, {Maruca}, {McComas}, {Messina}, {Mercer}, {Park}, {Peddie}, {Pogorelov}, {Reinhart}, {Richardson}, {Robinson}, {Rosen}, {Skoug}, {Slagle}, {Steinberg}, {Stevens}, {Szabo}, {Taylor}, {Tiu}, {Turin}, {Velli}, {Webb}, {Whittlesey}, {Wright}, {Wu}, and {Zank}]{kasper2016SSR}
J.~C. {Kasper}, R.~{Abiad}, G.~{Austin}, M.~{Balat-Pichelin}, S.~D. {Bale}, J.~W. {Belcher}, P.~{Berg}, H.~{Bergner}, M.~{Berthomier}, J.~{Bookbinder}, E.~{Brodu}, D.~{Caldwell}, A.~W. {Case}, B.~D.~G. {Chandran}, P.~{Cheimets}, J.~W. {Cirtain}, S.~R. {Cranmer}, D.~W. {Curtis}, P.~{Daigneau}, G.~{Dalton}, B.~{Dasgupta}, D.~{DeTomaso}, M.~{Diaz-Aguado}, B.~{Djordjevic}, B.~{Donaskowski}, M.~{Effinger}, V.~{Florinski}, N.~{Fox}, M.~{Freeman}, D.~{Gallagher}, S.~P. {Gary}, T.~{Gauron}, R.~{Gates}, M.~{Goldstein}, L.~{Golub}, D.~A. {Gordon}, R.~{Gurnee}, G.~{Guth}, J.~{Halekas}, K.~{Hatch}, J.~{Heerikuisen}, G.~{Ho}, Q.~{Hu}, G.~{Johnson}, S.~P. {Jordan}, K.~E. {Korreck}, D.~{Larson}, A.~J. {Lazarus}, G.~{Li}, R.~{Livi}, M.~{Ludlam}, M.~{Maksimovic}, J.~P. {McFadden}, W.~{Marchant}, B.~A. {Maruca}, D.~J. {McComas}, L.~{Messina}, T.~{Mercer}, S.~{Park}, A.~M. {Peddie}, N.~{Pogorelov}, M.~J. {Reinhart}, J.~D. {Richardson}, M.~{Robinson}, I.~{Rosen}, R.~M. {Skoug}, A.~{Slagle}, J.~T. {Steinberg}, M.~L. {Stevens},
  A.~{Szabo}, E.~R. {Taylor}, C.~{Tiu}, P.~{Turin}, M.~{Velli}, G.~{Webb}, P.~{Whittlesey}, K.~{Wright}, S.~T. {Wu}, and G.~{Zank}.
\newblock {Solar Wind Electrons Alphas and Protons (SWEAP) Investigation: Design of the Solar Wind and Coronal Plasma Instrument Suite for Solar Probe Plus}.
\newblock \emph{\ssr}, 204:\penalty0 131--186, December 2016.
\newblock \doi{10.1007/s11214-015-0206-3}.

\bibitem[{Kiyani} et~al.(2015){Kiyani}, {Osman}, and {Chapman}]{Kiyani2015RSPTA}
K.~H. {Kiyani}, K.~T. {Osman}, and S.~C. {Chapman}.
\newblock {Dissipation and heating in solar wind turbulence: from the macro to the micro and back again}.
\newblock \emph{Philosophical Transactions of the Royal Society of London Series A}, 373\penalty0 (2041):\penalty0 20140155--20140155, April 2015.
\newblock \doi{10.1098/rsta.2014.0155}.

\bibitem[{Kraichnan}(1965)]{kraichnan1965PoF}
R.~H. {Kraichnan}.
\newblock {Inertial-Range Spectrum of Hydromagnetic Turbulence}.
\newblock \emph{Physics of Fluids}, 8:\penalty0 1385--1387, July 1965.
\newblock \doi{10.1063/1.1761412}.

\bibitem[{Matthaeus} and {Goldstein}(1982)]{matthaeus1982JGR}
W.~H. {Matthaeus} and M.~L. {Goldstein}.
\newblock {Measurement of the rugged invariants of magnetohydrodynamic turbulence in the solar wind}.
\newblock \emph{\jgr}, 87:\penalty0 6011--6028, August 1982.
\newblock \doi{10.1029/JA087iA08p06011}.

\bibitem[{Matthaeus} and {Velli}(2011)]{matthaeus2011SSR}
W.~H. {Matthaeus} and M.~{Velli}.
\newblock {Who Needs Turbulence?. A Review of Turbulence Effects in the Heliosphere and on the Fundamental Process of Reconnection}.
\newblock \emph{\ssr}, 160:\penalty0 145--168, October 2011.
\newblock \doi{10.1007/s11214-011-9793-9}.

\bibitem[{Matthaeus} et~al.(1998){Matthaeus}, {Smith}, and {Oughton}]{matthaeus1998JGR}
W.~H. {Matthaeus}, C.~W. {Smith}, and S.~{Oughton}.
\newblock {Dynamical age of solar wind turbulence in the outer heliosphere}.
\newblock \emph{\jgr}, 103:\penalty0 6495, April 1998.
\newblock \doi{10.1029/97JA03729}.

\bibitem[{Matthaeus} et~al.(2004){Matthaeus}, {Minnie}, {Breech}, {Parhi}, {Bieber}, and {Oughton}]{matthaeus2004grl}
W.~H. {Matthaeus}, J.~{Minnie}, B.~{Breech}, S.~{Parhi}, J.~W. {Bieber}, and S.~{Oughton}.
\newblock {Transport of cross helicity and radial evolution of Alfv{\'e}nicity in the solar wind}.
\newblock \emph{\grl}, 31:\penalty0 L12803, June 2004.
\newblock \doi{10.1029/2004GL019645}.

\bibitem[{McComas} et~al.(2000){McComas}, {Barraclough}, {Funsten}, {Gosling}, {Santiago-Mu{\~n}oz}, {Skoug}, {Goldstein}, {Neugebauer}, {Riley}, and {Balogh}]{mccomas2000JGR}
D.~J. {McComas}, B.~L. {Barraclough}, H.~O. {Funsten}, J.~T. {Gosling}, E.~{Santiago-Mu{\~n}oz}, R.~M. {Skoug}, B.~E. {Goldstein}, M.~{Neugebauer}, P.~{Riley}, and A.~{Balogh}.
\newblock {Solar wind observations over Ulysses' first full polar orbit}.
\newblock \emph{\jgr}, 105:\penalty0 10419--10434, May 2000.
\newblock \doi{10.1029/1999JA000383}.

\bibitem[McComas et~al.(2025)McComas, Christian, Schwadron, Gkioulidou, Allegrini, Baker, Bzowski, Clark, Cohen, Cohen, et~al.]{MccomasEA25-IMAP}
David~J McComas, Eric~R Christian, NA~Schwadron, Matina Gkioulidou, F~Allegrini, DN~Baker, M~Bzowski, G~Clark, CMS Cohen, I~Cohen, et~al.
\newblock Interstellar mapping and acceleration probe: The nasa imap mission.
\newblock \emph{Space science reviews}, 221\penalty0 (8):\penalty0 100, 2025.

\bibitem[McComas et~al.(1998)McComas, Bame, Barker, Feldman, Phillips, Riley, and Griffee]{mccomas1998solar}
DJ~McComas, SJ~Bame, P~Barker, WC~Feldman, JL~Phillips, P~Riley, and JW~Griffee.
\newblock Solar wind electron proton alpha monitor (swepam) for the advanced composition explorer.
\newblock \emph{Space Science Reviews}, 86\penalty0 (1):\penalty0 563--612, 1998.

\bibitem[{McIntyre} et~al.(2023){McIntyre}, {Chen}, and {Larosa}]{McIntyre2023ApJ}
J.~R. {McIntyre}, C.~H.~K. {Chen}, and A.~{Larosa}.
\newblock {Properties Underlying the Variation of the Magnetic Field Spectral Index in the Inner Solar Wind}.
\newblock \emph{\apj}, 957\penalty0 (2):\penalty0 111, November 2023.
\newblock \doi{10.3847/1538-4357/acf3dd}.

\bibitem[{M{\"u}ller} et~al.(2020){M{\"u}ller}, {St. Cyr}, {Zouganelis}, {Gilbert}, {Marsden}, {Nieves-Chinchilla}, {Antonucci}, {Auch{\`e}re}, {Berghmans}, {Horbury}, {Howard}, {Krucker}, {Maksimovic}, {Owen}, {Rochus}, {Rodriguez-Pacheco}, {Romoli}, {Solanki}, {Bruno}, {Carlsson}, {Fludra}, {Harra}, {Hassler}, {Livi}, {Louarn}, {Peter}, {Sch{\"u}hle}, {Teriaca}, {del Toro Iniesta}, {Wimmer-Schweingruber}, {Marsch}, {Velli}, {De Groof}, {Walsh}, and {Williams}]{Muller2020AA}
D.~{M{\"u}ller}, O.~C. {St. Cyr}, I.~{Zouganelis}, H.~R. {Gilbert}, R.~{Marsden}, T.~{Nieves-Chinchilla}, E.~{Antonucci}, F.~{Auch{\`e}re}, D.~{Berghmans}, T.~S. {Horbury}, R.~A. {Howard}, S.~{Krucker}, M.~{Maksimovic}, C.~J. {Owen}, P.~{Rochus}, J.~{Rodriguez-Pacheco}, M.~{Romoli}, S.~K. {Solanki}, R.~{Bruno}, M.~{Carlsson}, A.~{Fludra}, L.~{Harra}, D.~M. {Hassler}, S.~{Livi}, P.~{Louarn}, H.~{Peter}, U.~{Sch{\"u}hle}, L.~{Teriaca}, J.~C. {del Toro Iniesta}, R.~F. {Wimmer-Schweingruber}, E.~{Marsch}, M.~{Velli}, A.~{De Groof}, A.~{Walsh}, and D.~{Williams}.
\newblock {The Solar Orbiter mission. Science overview}.
\newblock \emph{\aap}, 642:\penalty0 A1, October 2020.
\newblock \doi{10.1051/0004-6361/202038467}.

\bibitem[{Pope}(2000)]{pope2000book}
S.~B. {Pope}.
\newblock \emph{{Turbulent Flows}}.
\newblock Cambridge University Press, August 2000.

\bibitem[{Raouafi} et~al.(2023){Raouafi}, {Matteini}, {Squire}, {Badman}, {Velli}, {Klein}, {Chen}, {Matthaeus}, {Szabo}, {Linton}, {Allen}, {Szalay}, {Bruno}, {Decker}, {Akhavan-Tafti}, {Agapitov}, {Bale}, {Bandyopadhyay}, {Battams}, {Ber{\v{c}}i{\v{c}}}, {Bourouaine}, {Bowen}, {Cattell}, {Chandran}, {Chhiber}, {Cohen}, {D'Amicis}, {Giacalone}, {Hess}, {Howard}, {Horbury}, {Jagarlamudi}, {Joyce}, {Kasper}, {Kinnison}, {Laker}, {Liewer}, {Malaspina}, {Mann}, {McComas}, {Niembro-Hernandez}, {Nieves-Chinchilla}, {Panasenco}, {Pokorn{\'y}}, {Pusack}, {Pulupa}, {Perez}, {Riley}, {Rouillard}, {Shi}, {Stenborg}, {Tenerani}, {Verniero}, {Viall}, {Vourlidas}, {Wood}, {Woodham}, and {Woolley}]{Raouafi2023SSR}
N.~E. {Raouafi}, L.~{Matteini}, J.~{Squire}, S.~T. {Badman}, M.~{Velli}, K.~G. {Klein}, C.~H.~K. {Chen}, W.~H. {Matthaeus}, A.~{Szabo}, M.~{Linton}, R.~C. {Allen}, J.~R. {Szalay}, R.~{Bruno}, R.~B. {Decker}, M.~{Akhavan-Tafti}, O.~V. {Agapitov}, S.~D. {Bale}, R.~{Bandyopadhyay}, K.~{Battams}, L.~{Ber{\v{c}}i{\v{c}}}, S.~{Bourouaine}, T.~A. {Bowen}, C.~{Cattell}, B.~D.~G. {Chandran}, R.~{Chhiber}, C.~M.~S. {Cohen}, R.~{D'Amicis}, J.~{Giacalone}, P.~{Hess}, R.~A. {Howard}, T.~S. {Horbury}, V.~K. {Jagarlamudi}, C.~J. {Joyce}, J.~C. {Kasper}, J.~{Kinnison}, R.~{Laker}, P.~{Liewer}, D.~M. {Malaspina}, I.~{Mann}, D.~J. {McComas}, T.~{Niembro-Hernandez}, T.~{Nieves-Chinchilla}, O.~{Panasenco}, P.~{Pokorn{\'y}}, A.~{Pusack}, M.~{Pulupa}, J.~C. {Perez}, P.~{Riley}, A.~P. {Rouillard}, C.~{Shi}, G.~{Stenborg}, A.~{Tenerani}, J.~L. {Verniero}, N.~{Viall}, A.~{Vourlidas}, B.~E. {Wood}, L.~D. {Woodham}, and T.~{Woolley}.
\newblock {Parker Solar Probe: Four Years of Discoveries at Solar Cycle Minimum}.
\newblock \emph{\ssr}, 219\penalty0 (1):\penalty0 8, February 2023.
\newblock \doi{10.1007/s11214-023-00952-4}.

\bibitem[{Richardson} and {Cane}(2010)]{Richardson2010SoPh}
I.~G. {Richardson} and H.~V. {Cane}.
\newblock {Near-Earth Interplanetary Coronal Mass Ejections During Solar Cycle 23 (1996 - 2009): Catalog and Summary of Properties}.
\newblock \emph{\solphys}, 264\penalty0 (1):\penalty0 189--237, June 2010.
\newblock \doi{10.1007/s11207-010-9568-6}.

\bibitem[Richardson and Cane(2024)]{Richardson2024CMElist}
Ian Richardson and Hilary Cane.
\newblock {Near-Earth Interplanetary Coronal Mass Ejections Since January 1996}, 2024.
\newblock URL \url{https://doi.org/10.7910/DVN/C2MHTH}.

\bibitem[{Roberts} et~al.(1987){Roberts}, {Goldstein}, {Klein}, and {Matthaeus}]{roberts1987JGRb}
D.~A. {Roberts}, M.~L. {Goldstein}, L.~W. {Klein}, and W.~H. {Matthaeus}.
\newblock {Origin and evolution of fluctuations in the solar wind - HELIOS observations and Helios-Voyager comparisons}.
\newblock \emph{\jgr}, 92:\penalty0 12023--12035, November 1987.
\newblock \doi{10.1029/JA092iA11p12023}.

\bibitem[{Roberts} et~al.(1992){Roberts}, {Goldstein}, {Matthaeus}, and {Ghosh}]{roberts1992jgr}
D.~A. {Roberts}, M.~L. {Goldstein}, W.~H. {Matthaeus}, and S.~{Ghosh}.
\newblock {Velocity shear generation of solar wind turbulence}.
\newblock \emph{\jgr}, 97:\penalty0 17, November 1992.
\newblock \doi{10.1029/92JA01144}.

\bibitem[{Ruiz} et~al.(2014){Ruiz}, {Dasso}, {Matthaeus}, and {Weygand}]{Ruiz2014SoPh}
M.~E. {Ruiz}, S.~{Dasso}, W.~H. {Matthaeus}, and J.~M. {Weygand}.
\newblock {Characterization of the Turbulent Magnetic Integral Length in the Solar Wind: From 0.3 to 5 Astronomical Units}.
\newblock \emph{Solar Physics}, 289:\penalty0 3917--3933, October 2014.
\newblock \doi{10.1007/s11207-014-0531-9}.

\bibitem[{Schwenn}(2006)]{Schwenn2006SSR_wind_sources}
R.~{Schwenn}.
\newblock {Solar Wind Sources and Their Variations Over the Solar Cycle}.
\newblock \emph{\ssr}, 124\penalty0 (1-4):\penalty0 51--76, June 2006.
\newblock \doi{10.1007/s11214-006-9099-5}.

\bibitem[{Shi} et~al.(2023){Shi}, {Velli}, {Lionello}, {Sioulas}, {Huang}, {Halekas}, {Tenerani}, {R{\'e}ville}, {Dakeyo}, {Maksimovi{\'c}}, and {Bale}]{Shi2023ApJ}
Chen {Shi}, Marco {Velli}, Roberto {Lionello}, Nikos {Sioulas}, Zesen {Huang}, Jasper~S. {Halekas}, Anna {Tenerani}, Victor {R{\'e}ville}, Jean-Baptiste {Dakeyo}, Milan {Maksimovi{\'c}}, and Stuart~D. {Bale}.
\newblock {Proton and Electron Temperatures in the Solar Wind and Their Correlations with the Solar Wind Speed}.
\newblock \emph{\apj}, 944\penalty0 (1):\penalty0 82, February 2023.
\newblock \doi{10.3847/1538-4357/acb341}.

\bibitem[Smith et~al.(1998)Smith, L'Heureux, Ness, Acuna, Burlaga, and Scheifele]{smith1998ace}
Charles~W Smith, Jacques L'Heureux, Norman~F Ness, Mario~H Acuna, Leonard~F Burlaga, and John Scheifele.
\newblock The ace magnetic fields experiment.
\newblock \emph{Space Science Reviews}, 86\penalty0 (1):\penalty0 613--632, 1998.

\bibitem[{Taylor}(1938)]{taylor1938ProcRSL}
G.~I. {Taylor}.
\newblock {The Spectrum of Turbulence}.
\newblock \emph{Proceedings of the Royal Society of London Series A}, 164:\penalty0 476--490, February 1938.
\newblock \doi{10.1098/rspa.1938.0032}.

\bibitem[{Telloni} et~al.(2021){Telloni}, {Sorriso-Valvo}, {Woodham}, {Panasenco}, {Velli}, {Carbone}, {Zank}, {Bruno}, {Perrone}, {Nakanotani}, {Shi}, {D'Amicis}, {De Marco}, {Jagarlamudi}, {Steinvall}, {Marino}, {Adhikari}, {Zhao}, {Liang}, {Tenerani}, {Laker}, {Horbury}, {Bale}, {Pulupa}, {Malaspina}, {MacDowall}, {Goetz}, {de Wit}, {Harvey}, {Kasper}, {Korreck}, {Larson}, {Case}, {Stevens}, {Whittlesey}, {Livi}, {Owen}, {Livi}, {Louarn}, {Antonucci}, {Romoli}, {O'Brien}, {Evans}, and {Angelini}]{Telloni2021ApJ}
Daniele {Telloni}, Luca {Sorriso-Valvo}, Lloyd~D. {Woodham}, Olga {Panasenco}, Marco {Velli}, Francesco {Carbone}, Gary~P. {Zank}, Roberto {Bruno}, Denise {Perrone}, Masaru {Nakanotani}, Chen {Shi}, Raffaella {D'Amicis}, Rossana {De Marco}, Vamsee~K. {Jagarlamudi}, Konrad {Steinvall}, Raffaele {Marino}, Laxman {Adhikari}, Lingling {Zhao}, Haoming {Liang}, Anna {Tenerani}, Ronan {Laker}, Timothy~S. {Horbury}, Stuart~D. {Bale}, Marc {Pulupa}, David~M. {Malaspina}, Robert~J. {MacDowall}, Keith {Goetz}, Thierry~Dudok {de Wit}, Peter~R. {Harvey}, Justin~C. {Kasper}, Kelly~E. {Korreck}, Davin {Larson}, Anthony~W. {Case}, Michael~L. {Stevens}, Phyllis {Whittlesey}, Roberto {Livi}, Christopher~J. {Owen}, Stefano {Livi}, Philippe {Louarn}, Ester {Antonucci}, Marco {Romoli}, Helen {O'Brien}, Vincent {Evans}, and Virginia {Angelini}.
\newblock {Evolution of Solar Wind Turbulence from 0.1 to 1 au during the First Parker Solar Probe-Solar Orbiter Radial Alignment}.
\newblock \emph{\apjl}, 912\penalty0 (2):\penalty0 L21, May 2021.
\newblock \doi{10.3847/2041-8213/abf7d1}.

\bibitem[{Usmanov} et~al.(2018){Usmanov}, {Matthaeus}, {Goldstein}, and {Chhiber}]{usmanov2018}
A.~V. {Usmanov}, W.~H. {Matthaeus}, M.~L. {Goldstein}, and R.~{Chhiber}.
\newblock {The Steady Global Corona and Solar Wind: A Three-dimensional MHD Simulation with Turbulence Transport and Heating}.
\newblock \emph{\apj}, 865:\penalty0 25, September 2018.
\newblock \doi{10.3847/1538-4357/aad687}.

\bibitem[{Usmanov} et~al.(2025){Usmanov}, {Chhiber}, {Matthaeus}, {Roy}, and {Goldstein}]{Usmanov2025ApJ}
Arcadi~V. {Usmanov}, Rohit {Chhiber}, William~H. {Matthaeus}, Sohom {Roy}, and Melvyn~L. {Goldstein}.
\newblock {A Unified Three-dimensional Magnetohydrodynamic Model of the Solar Corona, Solar Wind, and Global Heliosphere with Turbulence Transport}.
\newblock \emph{\apj}, 993\penalty0 (1):\penalty0 87, November 2025.
\newblock \doi{10.3847/1538-4357/ae019c}.

\bibitem[{Verscharen} et~al.(2019){Verscharen}, {Klein}, and {Maruca}]{Verscharen2019LRSP}
Daniel {Verscharen}, Kristopher~G. {Klein}, and Bennett~A. {Maruca}.
\newblock {The multi-scale nature of the solar wind}.
\newblock \emph{Living Reviews in Solar Physics}, 16\penalty0 (1):\penalty0 5, December 2019.
\newblock \doi{10.1007/s41116-019-0021-0}.

\bibitem[{Wang} et~al.(2024){Wang}, {Chhiber}, {Roy}, {Cuesta}, {Pecora}, {Yang}, {Fu}, {Li}, and {Matthaeus}]{Wang2024ApJ}
Jiaming {Wang}, Rohit {Chhiber}, Sohom {Roy}, Manuel~E. {Cuesta}, Francesco {Pecora}, Yan {Yang}, Xiangrong {Fu}, Hui {Li}, and William~H. {Matthaeus}.
\newblock {Anisotropy of Density Fluctuations in the Solar Wind at 1 au}.
\newblock \emph{\apj}, 967\penalty0 (2):\penalty0 150, June 2024.
\newblock \doi{10.3847/1538-4357/ad3e7a}.

\bibitem[{Zhou} et~al.(2004){Zhou}, {Matthaeus}, and {Dmitruk}]{zhou2004RMP}
Y.~{Zhou}, W.~H. {Matthaeus}, and P.~{Dmitruk}.
\newblock {Colloquium: Magnetohydrodynamic turbulence and time scales in astrophysical and space plasmas}.
\newblock \emph{Reviews of Modern Physics}, 76:\penalty0 1015--1035, December 2004.
\newblock \doi{10.1103/RevModPhys.76.1015}.

\end{thebibliography}

\end{document}